\documentclass[english,prb, preprint]{revtex4-1}
\usepackage[T1]{fontenc}
\usepackage[latin9]{inputenc}
\usepackage{color}
\usepackage{units}
\usepackage{amsmath}
\usepackage{graphicx}

\makeatletter

\providecommand{\tabularnewline}{\\}
\newcommand{\lyxdot}{.}

\@ifundefined{showcaptionsetup}{}{%
 \PassOptionsToPackage{caption=false}{subfig}}
\usepackage{subfig}
\makeatother

\usepackage{babel}
\begin{document}
\title{Graphene quantum dots with Stone-Wales defect as a topologically tunable
platform for visible-light harvesting }
\author{Tista Basak}
\address{Mukesh Patel School of Technology Management and Engineering, NMIMS
University, Mumbai 400056, India}
\email{Tista.Basak@nmims.edu}

\author{Tushima Basak }
\address{Department of Physics, Mithibai College, Mumbai 400056, India}
\email{Tushima.Basak@mithibai.ac.in}

\author{Alok Shukla}
\address{Department of Physics, Indian Institute of Technology Bombay, Powai,
Mumbai 400076, India}
\email{shukla@phy.iitb.ac.in}

\begin{abstract}
In this work, we report for the first time the crucial role of topological
anomalies like Stone-Wales (SW) type bond rotations in tuning the
optical properties of graphene quantum dots (GQDs). By means of first-principles
calculations, we first show that the structural stability of GQDs
strongly depends on position of SW defects. Optical absorption spectra
is then computed using electron-correlated methodology to demonstrate
that SW type reconstruction is responsible for the appearance of new
defect-induced peaks below the optical gap and dramatically modifies
the optical absorption profile. In addition, our investigations signify
that electron correlation effects become more dominant for SW-defected
GQDs. We finally establish that the introduction of SW defects at
specific locations strongly enhances light absorption in visible range,
which is of prime importance for designing light harvesting, photocatalytic
and optoelectronic devices.
\end{abstract}
\maketitle

\section{Introduction}

Currently, the contribution of solar energy to the total global energy
demand is quite low and consequently tremendous scientific efforts
have been directed to efficiently design devices for visible light
harvesting. In recent years, graphene quantum dots have emerged as
a promising solution for developing different optoelectronic devices,
bio-imaging sensors, etc. due to their unique electronic and tunable
optical properties \citep{Wang}.\textcolor{red}{{} }Despite this recent
progress, \textcolor{black}{very little literature exists} on the
potential application of GQDs in visible light-sensitive devices since
most of the GQDs are efficient absorbers of light in the UV range,
rather than the visible range. In this work, we demonstrate for the
first time that systematic introduction of topological disorders such
as Stone-Wales defects in GQDs in a controlled manner can significantly
enhance light absorption in the visible range, opening up their prospective
applications in the field of photovoltaics. 

The SW defect formed due to rotation of carbon-carbon bonds with respect
to their midpoint by 90$^{0}$, leading to the transformation of four
hexagons into two pentagons and two heptagons, has attracted substantial
attention as it has the lowest formation energy among the different
types of topological anomalies in graphenic systems. Experimental
investigations have demonstrated the existence of SW-type bond rotations
along the edge \citep{He_ACS_Nano_2015,Girit_Science_2009,Chuvilin_2009_New_J_Phys}
and also in the core \citep{Meyer_Nano_Letters_2008} of graphene
and its nano-structures. The presence of SW defects along the zigzag
boundary leads to the formation of a reconstructed zigzag edge (reczag
edge) which is experimentally confirmed to be stable in single-walled
carbon nanotube for a time span of several seconds \citep{Koskinen_PRB_2009}.
\textcolor{black}{The sequential reconstruction of zigzag edge of
GNRs into reczag edge as a function of time and temperature has also
been investigated theoretically} \citep{Lee_PRB_2010}. Theoretical
studies indicate that energy stability of different edge configurations
is a highly debated issue, with critical dependence on the dimension
of graphene nanostructures and concentration of hydrogen atoms \citep{Koskinen_PRL_2008,Wassmann_PRL_2008,Voznyy_PRB_2011}.\textbf{
}Inspite of availability of\textbf{ }substantial amount of literature
to demonstrate the existence of SW defects and its consequence on
the electronic, magnetic and mechanical properties of graphene nanostructures
\citep{Bhowmick_PRB_2010,Romanovsky_PRB_2012,Rodrigues_PRB_2011,Ihnatsenka_PRB_2013},
no studies have been initiated till date to the best of our knowledge,
to explore the influence of these defects on the optical properties
of these systems. 

\textcolor{black}{In the present work, we have studied the structure,
energetics, and optical properties of the SW type reconstructions
in a reasonably large hydrogen-passivated graphene quantum dot with
64 carbon atoms (GQD-64). We considered two locations for the SW defect:
(i) one of the zigzag edges resulting in a reczag edge (SW1-GQD-64),
and (ii) core of GQD-64 (SW2-GQD-64), and employed a first-principles
methodology to probe their energetics. For computing their optical
properties, we used a computational approach based on the Pariser-Parr-Pople
(PPP) model Hamiltonian, coupled with the configuration-interaction
(CI) methodology for including the electron-correlation effects. Our
studies reveal that the introduction of SW-type defects in GQDs significantly
alters their electronic structure and optical properties as compared
to those of the pristine system. Furthermore, our results demonstrate
that the striking attributes of these properties are critically dependent
upon the location of SW-type defects. }

\textcolor{black}{The rest of the paper is organized as follows. The
computational methodology employed by us is briefly described in Sec.
\ref{sec:Computational-Methodology} which is followed by the results
and discussion in Sec. \ref{sec:Results-and-Discussion}. Finally,
the conclusions are presented in Sec. \ref{sec:Conclusions}.}

\section{Computational \label{sec:Computational-Methodology}Methodology}

\begin{figure}
\subfloat[]{\includegraphics[scale=0.14]{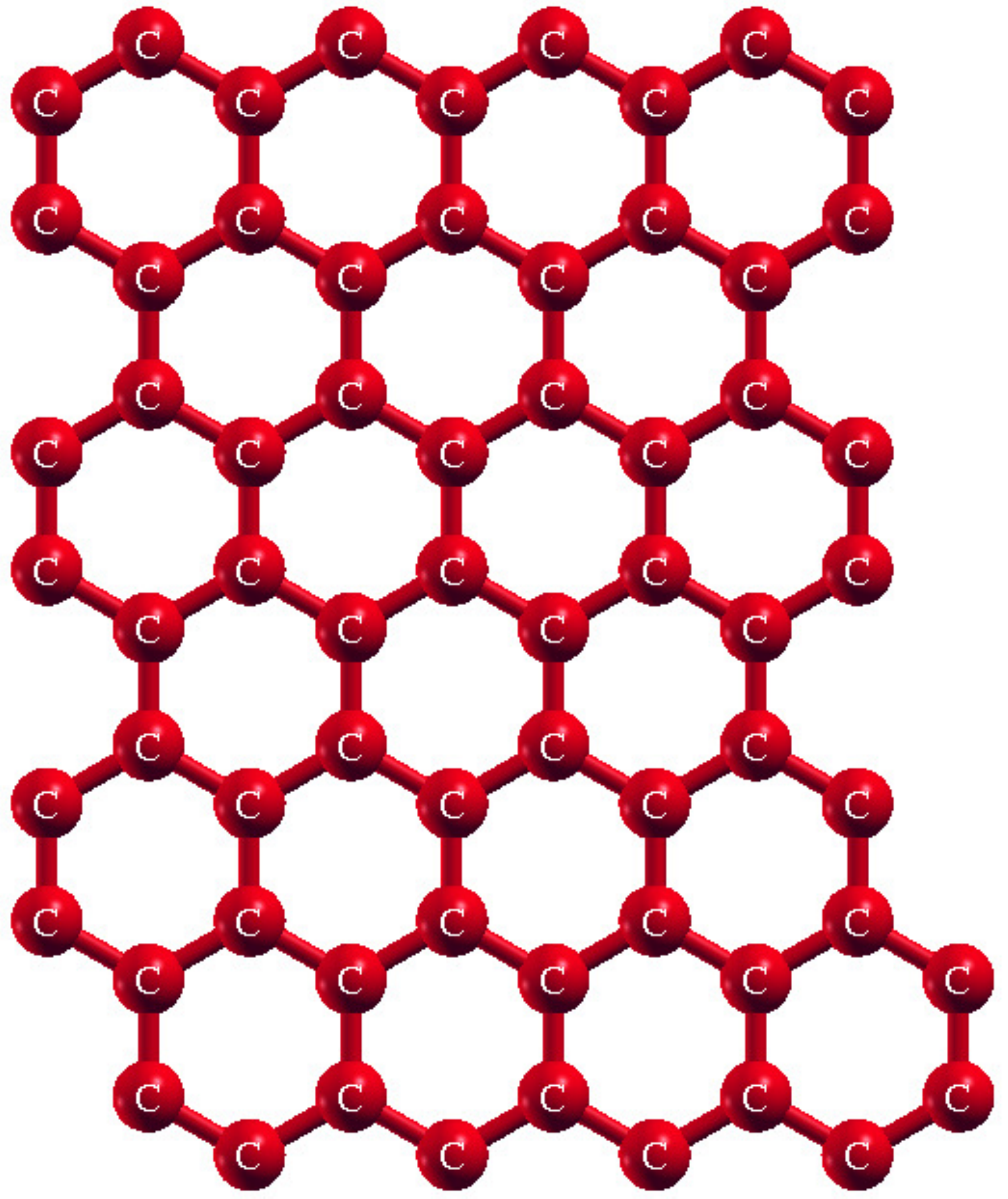}

}\subfloat[]{\includegraphics[scale=0.15]{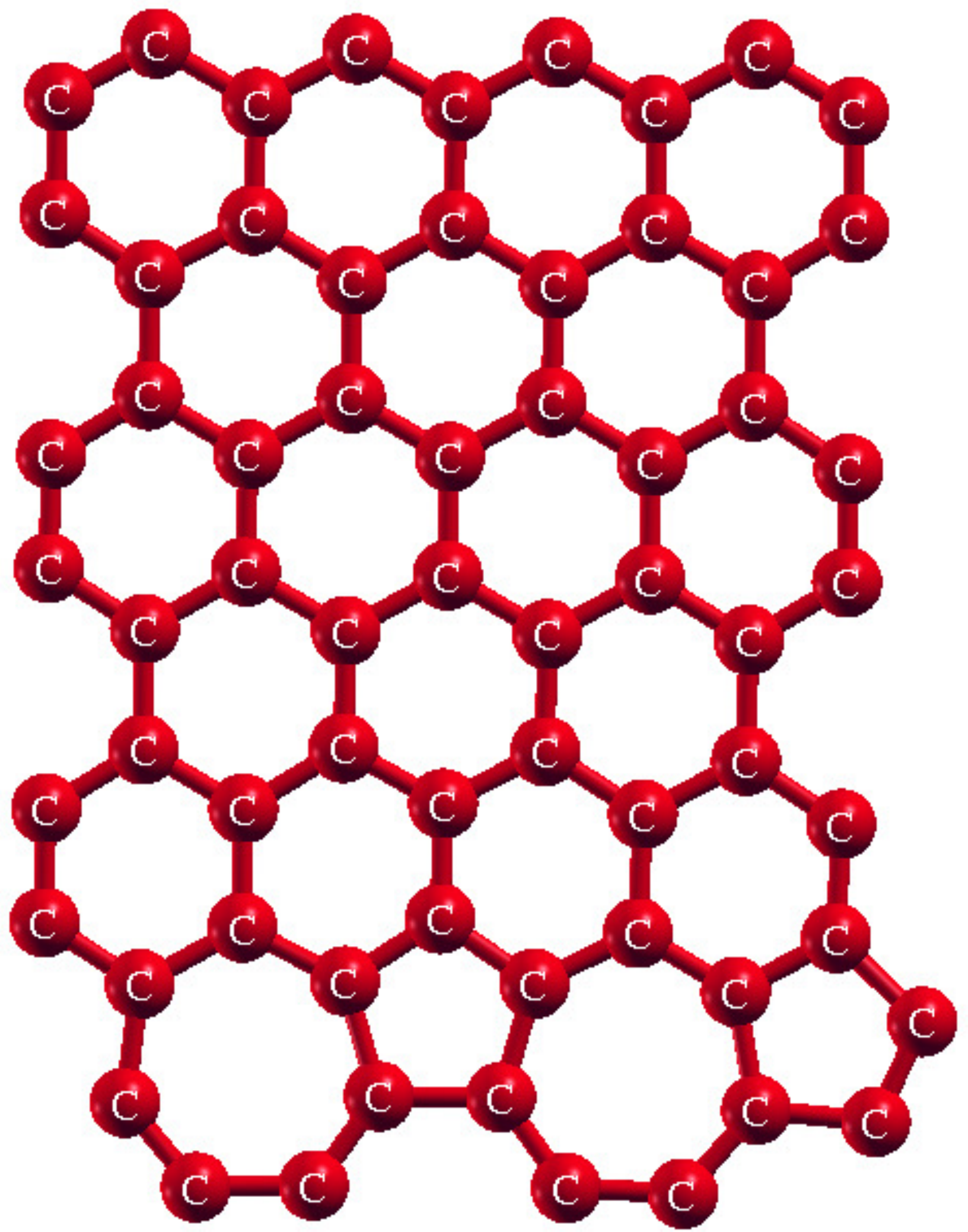}

}\subfloat[]{\includegraphics[scale=0.15]{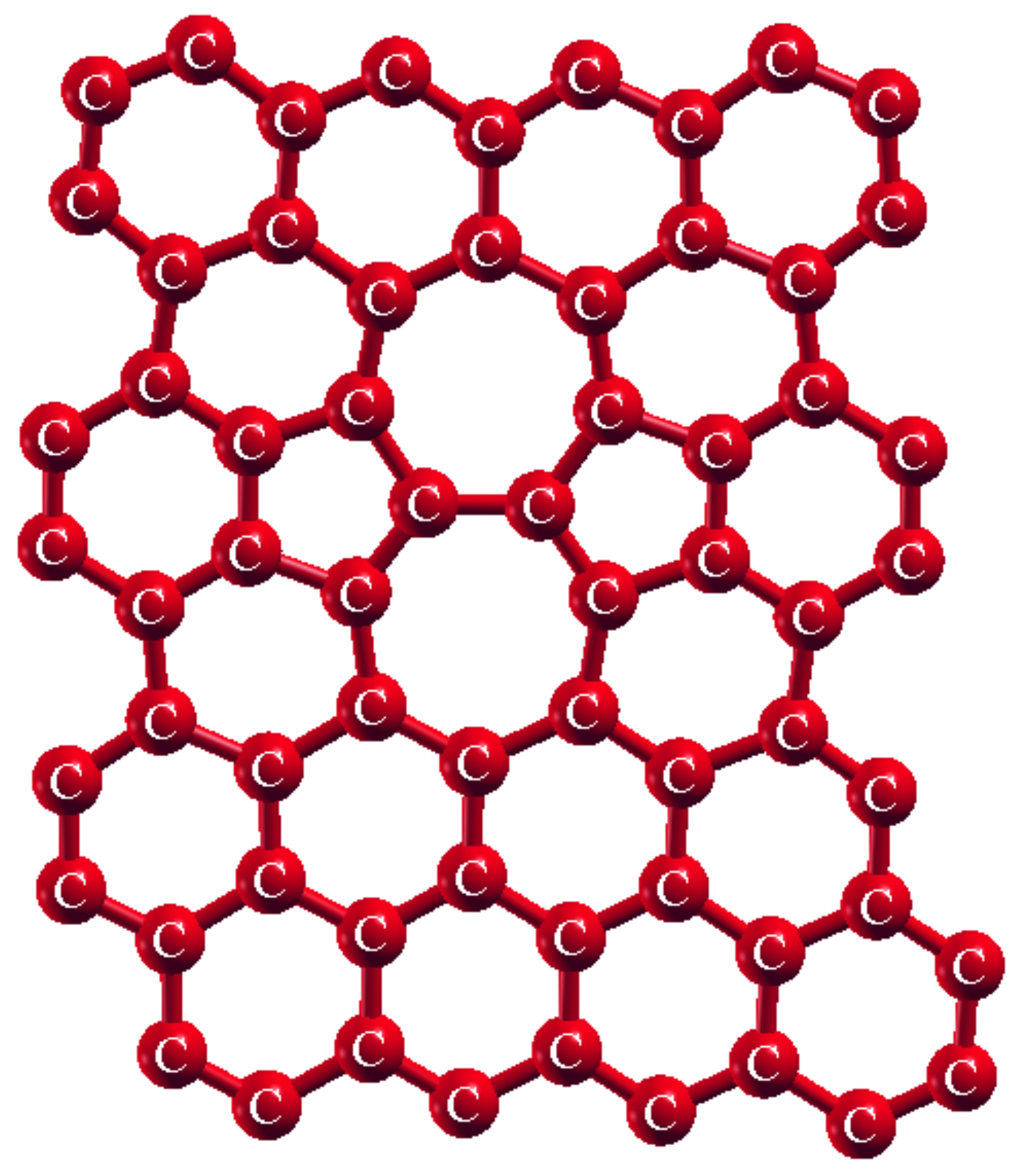}

}\caption{\textcolor{black}{\label{fig:Schematic-representation-of}Schematic
representation of hydrogen passivated (a) GQD-64, }(b) SW1-GQD-64\textcolor{black}{{}
and }(c) SW2-GQD-64.}

\end{figure}

\textcolor{black}{The schematic representations of hydrogen-passivated
GQD-64, SW1-GQD-64 and SW2-GQD-64, lying in the x-y plane, are given
in Fig. \ref{fig:Schematic-representation-of}. The bond-lengths and
bond-angles between carbon atoms are chosen uniformly as 1.4 Å and
$120^{0}$, respectively, for GQD-64. In case of SW1-GQD-64 and SW2-GQD-64
configurations, the optimized geometries obtained with Gaussian 16
program package\citep{g16} using cc-pvdz basis set at restricted
Hartree-Fock (RHF) level, exhibit non-uniform C-C bond-lengths in
the range (1.35 - 1.49) $\textrm{Å}$. The correlated computations
on these relaxed geometries have been performed by employing our theoretical
formalism based upon the effective $\pi$-electron Pariser-Parr-Pople
(PPP) model Hamiltonian \citep{ppp-pople,ppp-pariser-parr}}

\textcolor{black}{
\begin{align}
\mbox{\mbox{\mbox{\mbox{\ensuremath{H}}}}} & \ensuremath{=}\ensuremath{-}\ensuremath{\sum_{i,j,\sigma}t_{ij}\left(c_{i\sigma}^{\dagger}c_{j\sigma}+c_{j\sigma}^{\dagger}c_{i\sigma}\right)}\ensuremath{+}\ensuremath{U\sum_{i}n_{i\uparrow}n_{i\downarrow}}\nonumber \\
 & \ensuremath{+}\ensuremath{\sum_{i<j}V_{ij}(n_{i}-1)(n_{j}-1)}\label{eq:ppp}
\end{align}
}

\textcolor{black}{where a $\pi$ orbital of spin $\sigma$, localized
on the ith carbon atom is created (annihilated) by $c_{i\sigma}^{\dagger}($c$_{i\sigma})$
while $n$$_{i}=\sum_{\sigma}c_{i\sigma}^{\dagger}c_{i\sigma}$ denotes
the total number of $\pi$-electrons with spin $\sigma$ on atom $i$.
The electron-electron repulsion is included by the second and the
third terms in Eq. \ref{eq:ppp}, with the parameters $U$ and $V_{ij}$
representing the on-site and long-range Coulomb interactions, respectively.
The one-electron hopping matrix elements $t_{ij}$ are restricted
between nearest-neighbor carbon atoms $i$ and $j$, with the value
$t_{0}$= -2.4 eV corresponding to uniform carbon-carbon bond-length
$r_{0}=1.4$ Å, in accordance with our earlier calculations on conjugated
polymers, polyaromatic hydrocarbons, and graphene quantum dots\citep{PhysRevB.65.125204Shukla,PhysRevB.69.165218Shukla69,Tista_PRB92,Tista_PRB93,PhysRevB.98.035401,Tista_Basak_Mat_Today_Proc_2020,Tushima_Basak_Mat_Today_Proc_2020}.
For the non-uniform bond-lengths, the values of corresponding $t_{ij}$
are determined from the exponential formula $t_{ij}=t_{0}e^{(r_{0}-r_{ij})/\delta}$,
extensively used by us earlier\citep{DKRai_Hopping}, in which $r_{ij}$
is the distance between $i$th and $j$th carbon atoms (in $\textrm{Å}$),
$t_{0}$= -2.4 eV $r_{0}=1.4$ Å, and $\delta=0.73$ Å is a parameter
depicting electron-phonon coupling. }

\textcolor{black}{The long-range Coulomb interactions, $V_{ij}$ are
parametrized as per the Ohno relationship\citep{Theor.chim.act.2Ohno} }

\textcolor{black}{
\begin{equation}
V_{ij}=U/\kappa_{i,j}(1+0.6117R_{i,j}^{2})^{\nicefrac{1}{2}}
\end{equation}
}

\textcolor{black}{where $\kappa_{i,j}$ is the dielectric constant
of the system representing the screening effects, and $R_{i,j}$ is
the distance between $i$th and $j$th carbon atoms (in $\textrm{Å}$).
In the present set of computations, we have adopted the ``screened
parameters''\citep{PhysRevB.55.1497Chandross} with $U=8.0$ eV,
$\kappa_{i,j}=2.0\:(i\neq j)$, and $\kappa_{i,i}=1.0$, consistent
with our earlier works on $\pi$-conjugated systems and graphene quantum
dots\citep{PhysRevB.65.125204Shukla,PhysRevB.69.165218Shukla69,Tista_PRB92,Tista_PRB93,PhysRevB.98.035401,Tista_Basak_Mat_Today_Proc_2020,Tushima_Basak_Mat_Today_Proc_2020}.}

\textcolor{black}{Our computations are initiated by applying the mean-field
approximation at the RHF level with the PPP Hamiltonian (Eq. \ref{eq:ppp}),
using a code developed in our group\citep{Sony2010821}. The molecular
orbitals (MOs) derived from the RHF calculations are employed to transform
the PPP Hamiltonian from the site basis to MO basis for incorporating
electron correlations by the configuration interaction (CI) approach.
The CI calculations are performed at the multi-reference singles-doubles
configuration interaction (MRSDCI) level in which, single and double
excitations from an initial reference space consisting of closed-shell
RHF ground state are employed for computing the CI matrix. The eigenfunctions
obtained from the CI matrix are used to calculate transition electric
dipole matrix elements between various states, essential for computing
the optical absorption spectra. The excited states responsible for
various peaks in the absorption spectra are determined, and their
corresponding reference configurations having coefficients above a
chosen convergence threshold are included to augment the reference
space. This enhanced reference set is then utilized to perform the
next MRSDCI calculation. The above sequence of operations is iterated
until the excitation energies and optical absorption spectra of the
system converge to an acceptable tolerance. In order to reduce the
size of the CI expansion matrix, and make the computations feasible,
the frozen orbital approximation is adopted, wherein a few of the
lowest energy occupied orbitals are frozen, and high-energy virtual
orbitals are deleted, leading to a smaller set of active MOs.}

\textcolor{black}{The ground state optical absorption cross-section
$\sigma(\omega)$ assuming a Lorentzian line shape is computed using
the transition dipole matrix elements, and the energies of the excited
states, according to the formula
\begin{equation}
\sigma(\omega)=4\pi\alpha\underset{i}{\sum}\frac{\omega_{i0}\left|\left\langle i\left|\mathbf{\hat{e}.r}\right|0\right\rangle \right|^{2}\gamma}{\left(\omega_{i0}-\omega\right)^{2}+\gamma^{2}}
\end{equation}
}

\textcolor{black}{where $\omega$ represents the frequency of incident
radiation, $\hat{{\bf e}}$ denotes its polarization direction, ${\bf r}$
is the position operator, $\omega_{i0}$ is the frequency difference
between $0$ (ground) and $i$ (excited) states, $\alpha$ is the
fine structure constant, and $\gamma$ is the absorption line-width.}

\section{Results \label{sec:Results-and-Discussion}and Discussion}

In this section, we will first investigate the energy stability, electronic
band gaps, and the charge density plots of GQD-64, SW1-GQD-64 and
SW2-GQD-64. Thereafter, the conspicuous role played by electron correlation
effects in governing the optical properties of these systems are emphasized
by a critical analysis of the results obtained from independent particle
(TB), PPP model at the restricted Hartree-Fock level (PPP-HF), and
the configuration-interaction (CI) methodology (PPP-CI).

\subsection{Energy stability, energy band-gap and charge density plots}

In Table \textcolor{black}{\ref{tab:Energy-band-gap-obtained}} we
report (i) the difference in energy ($\Delta E$) of optimized SW1-GQD-64
and SW2-GQD-64 configurations with respect to pristine GQD-64 (computed
from Gaussian 16 program package) and (ii) energy band-gap between
the highest occupied (H) and lowest unoccupied (L) molecular orbitals
obtained from TB, PPP-HF and the correlated calculations using the
PPP-CI approach. A comparison of the energetics of these three systems
reveal that GQD-64 has lowest energy followed by SW1-GQD-64 and SW2-GQD-64.
This indicates that reconstruction of zigzag to reczag edge (SW1-GQD-64)
increases the energy of the system in agreement with earlier theoretical
predictions on hydrogen passivated triangular-shaped GQD \citep{Voznyy_PRB_2011}.
Further, quantum dots with reczag edge are more stable as compared
to structures having SW defect at the middle (SW2-GQD-64). This implies
that the energy stability of GQDs depends crucially on the location
of SW defects. At TB and PPP-HF levels, the computed H-L gaps for
GQD-64 and SW2-GQD-64 are very close, while the gap is comparatively
higher for SW1-GQD-64 configuration. However, on performing the electron-correlated
PPP-CI calculations, the gaps for GQD-64 and SW2-GQD-64 increase as
compared to their PPP-HF values, while that of SW1-GQD-64 decreases
slightly. In addition, these band-gap values are significantly higher
than independent particle model results, signifying the importance
of electron repulsion effects in these systems. Furthermore, discernible
deviation in the gaps calculated at the HF and the CI levels for GQD-64
and SW2-GQD-64, in contrast to their almost identical values for SW1-GQD-64,
imply stronger electron-correlation effects in GQD-64 and SW2-GQD-64
as compared to SW1-GQD-64. 

\begin{table}
\caption{\label{tab:Energy-band-gap-obtained}Energy difference ($\Delta E)$
of SW defected structures with respect to pristine GQD-64 and their
\textcolor{black}{HOMO (H)-LUMO (L)} band-gap obtained from TB, PPP
model at the RHF level and the PPP-CI calculations. }
\begin{tabular}{|c|c|c|c|c|}
\hline 
 & $\Delta E$ & \multicolumn{3}{c|}{H-L band-gap (eV)}\tabularnewline
\cline{3-5} \cline{4-5} \cline{5-5} 
System & (eV) & TB & PPP-HF & PPP-CI\tabularnewline
\hline 
GQD-64 & 0 & 0.13 & 0.74 & 1.15\tabularnewline
\hline 
SW1-GQD-64 & 2.77 & 0.32 & 0.89 & 0.87\tabularnewline
\hline 
SW2-GQD-64 & 4.14 & 0.14 & 0.72 & 1.05\tabularnewline
\hline 
\end{tabular}
\end{table}

\begin{figure}
\subfloat[]{\includegraphics[width=2.5cm]{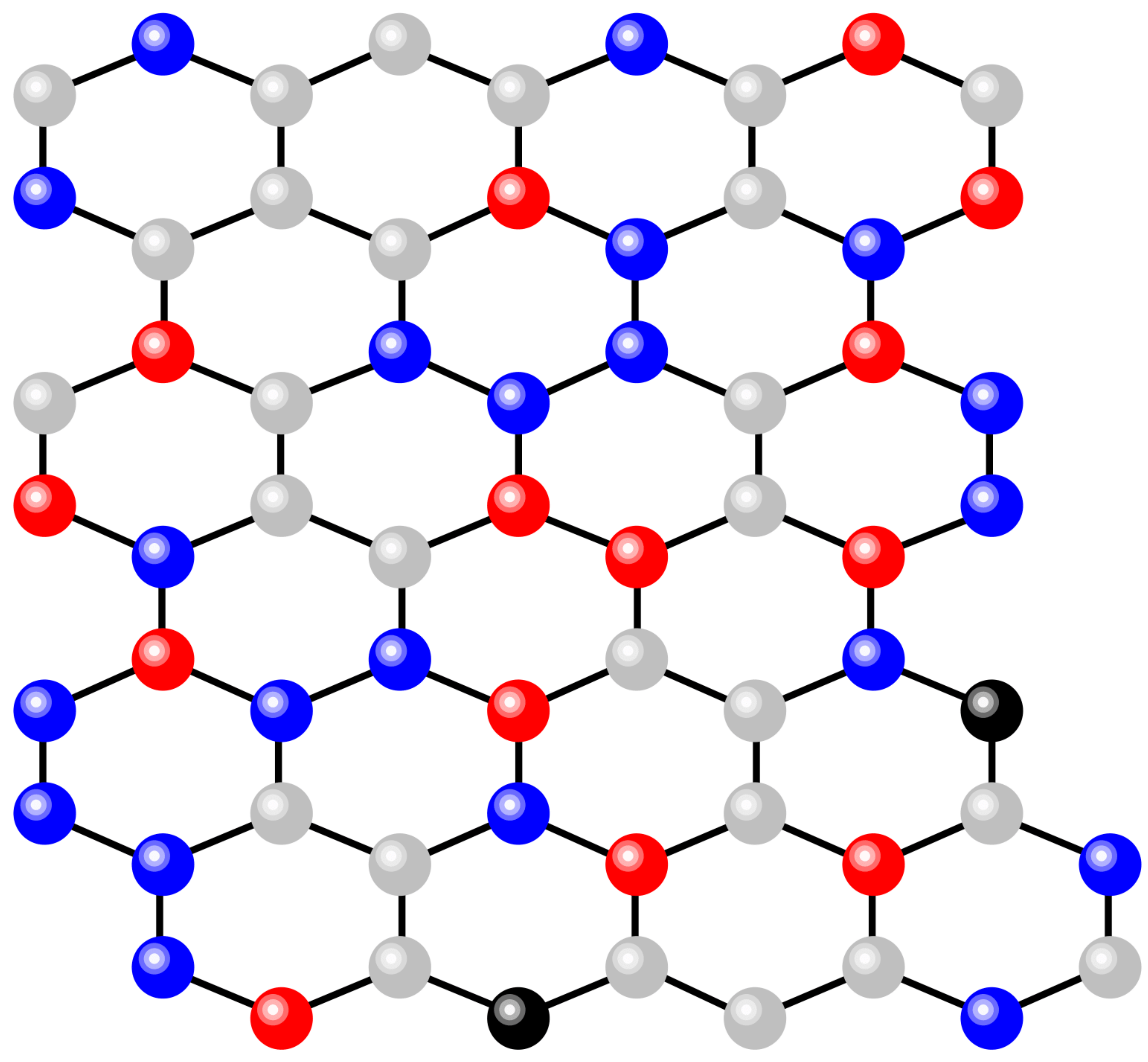}}\subfloat[]{\includegraphics[width=2.5cm]{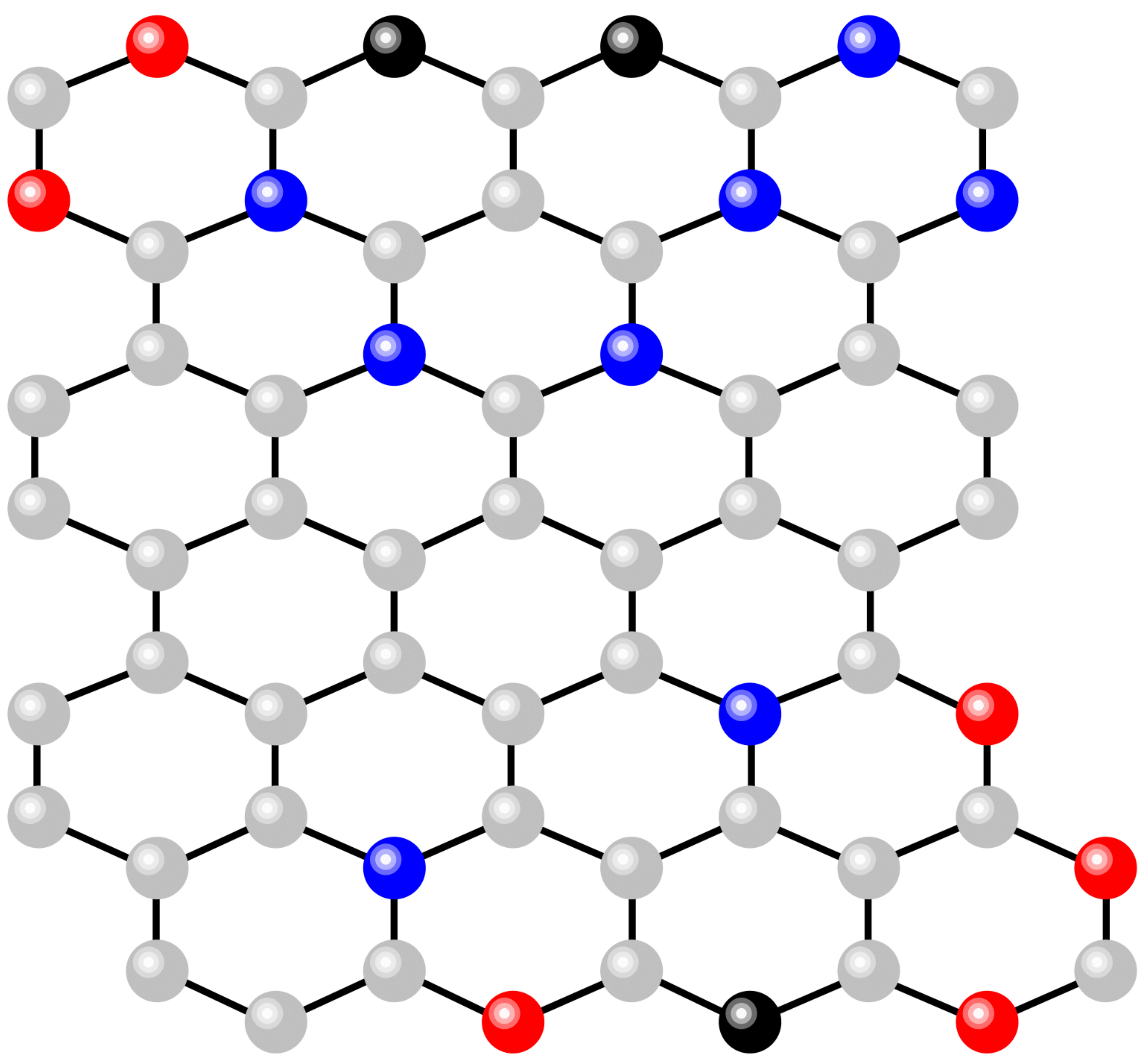}}\subfloat[]{\includegraphics[width=2.5cm]{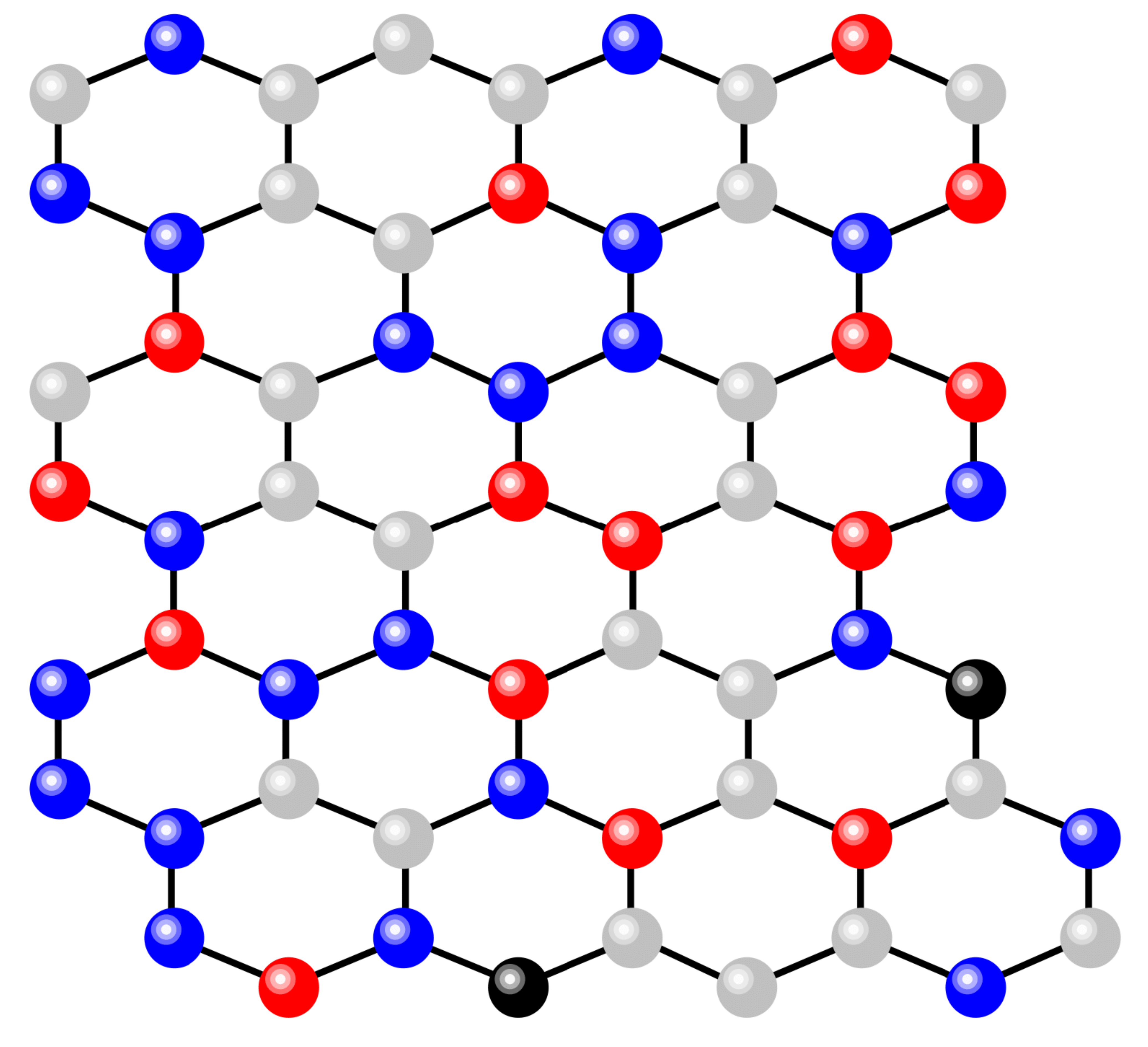}}\subfloat[]{\includegraphics[width=2.5cm]{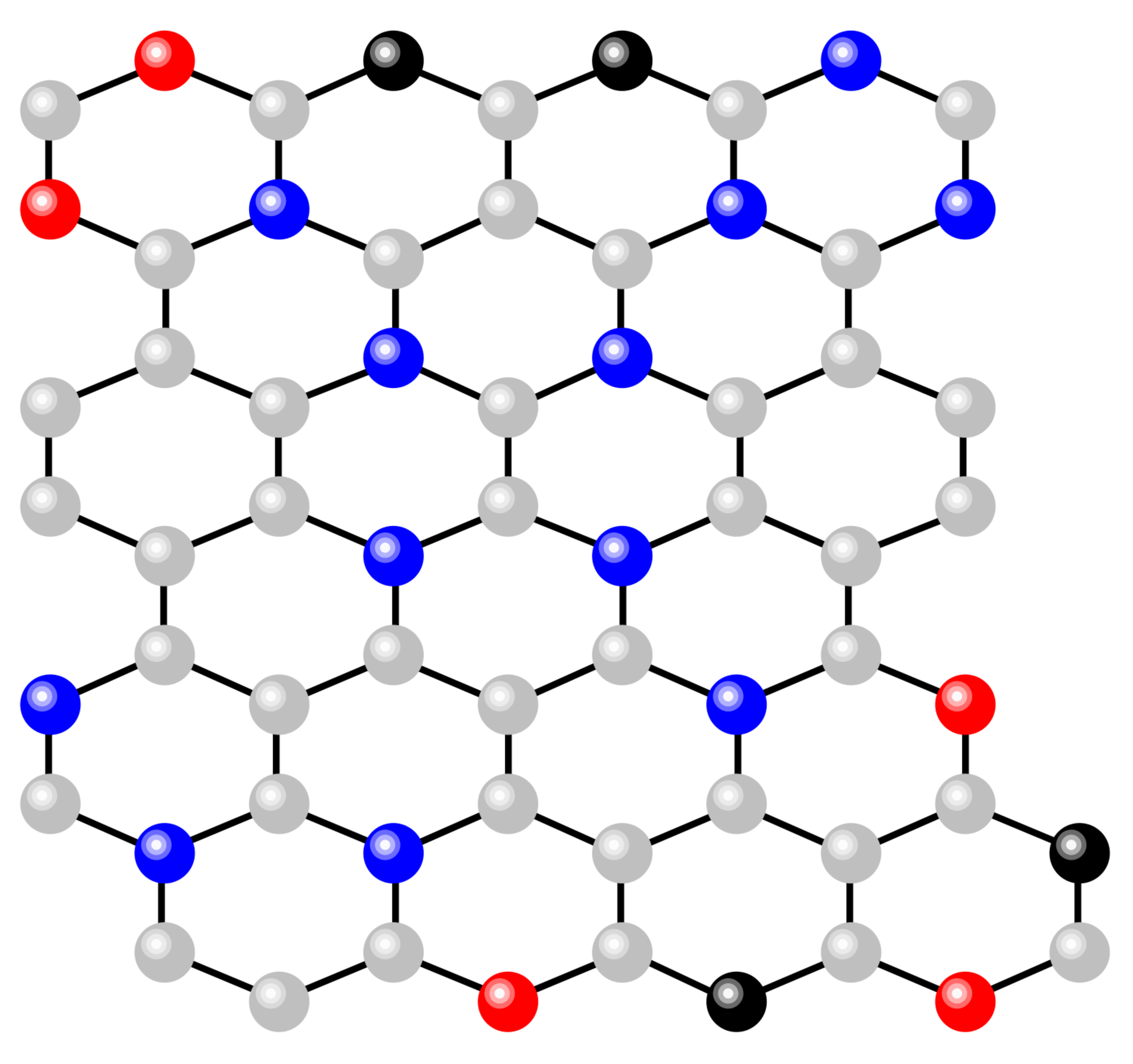}}

\subfloat[]{\includegraphics[width=2.2cm]{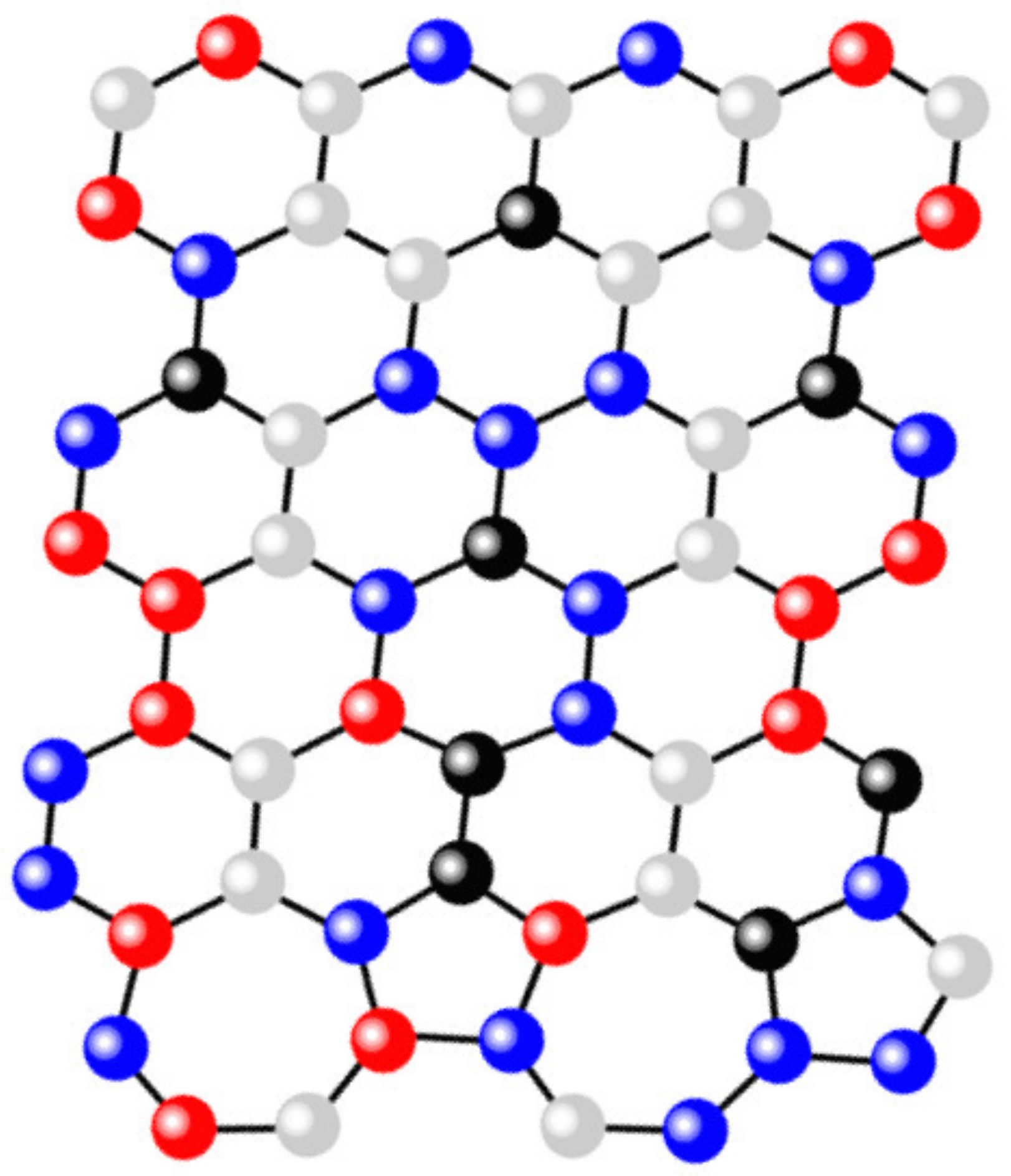}}\subfloat[]{\includegraphics[width=2.2cm]{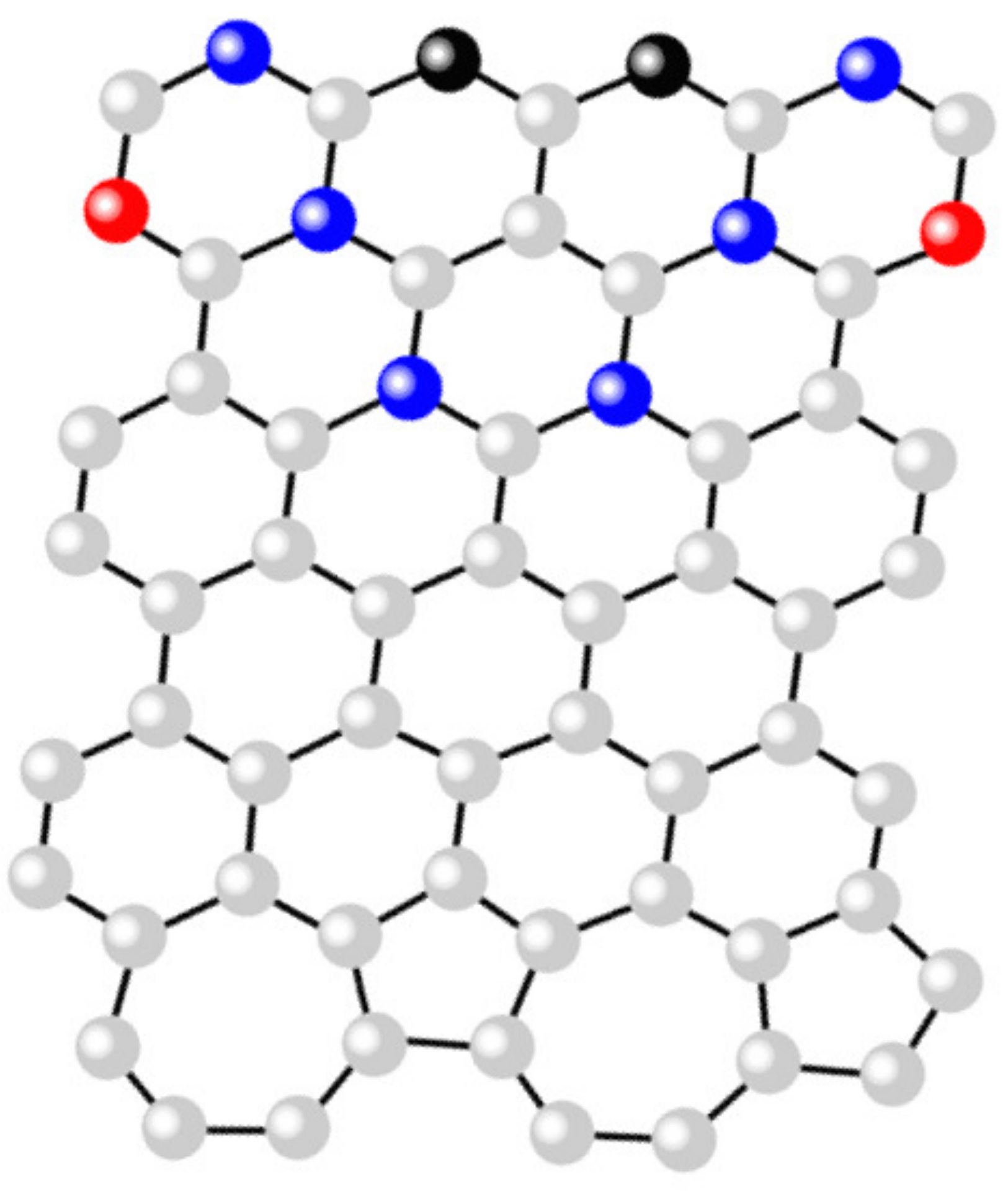}}\subfloat[]{\includegraphics[width=2.2cm]{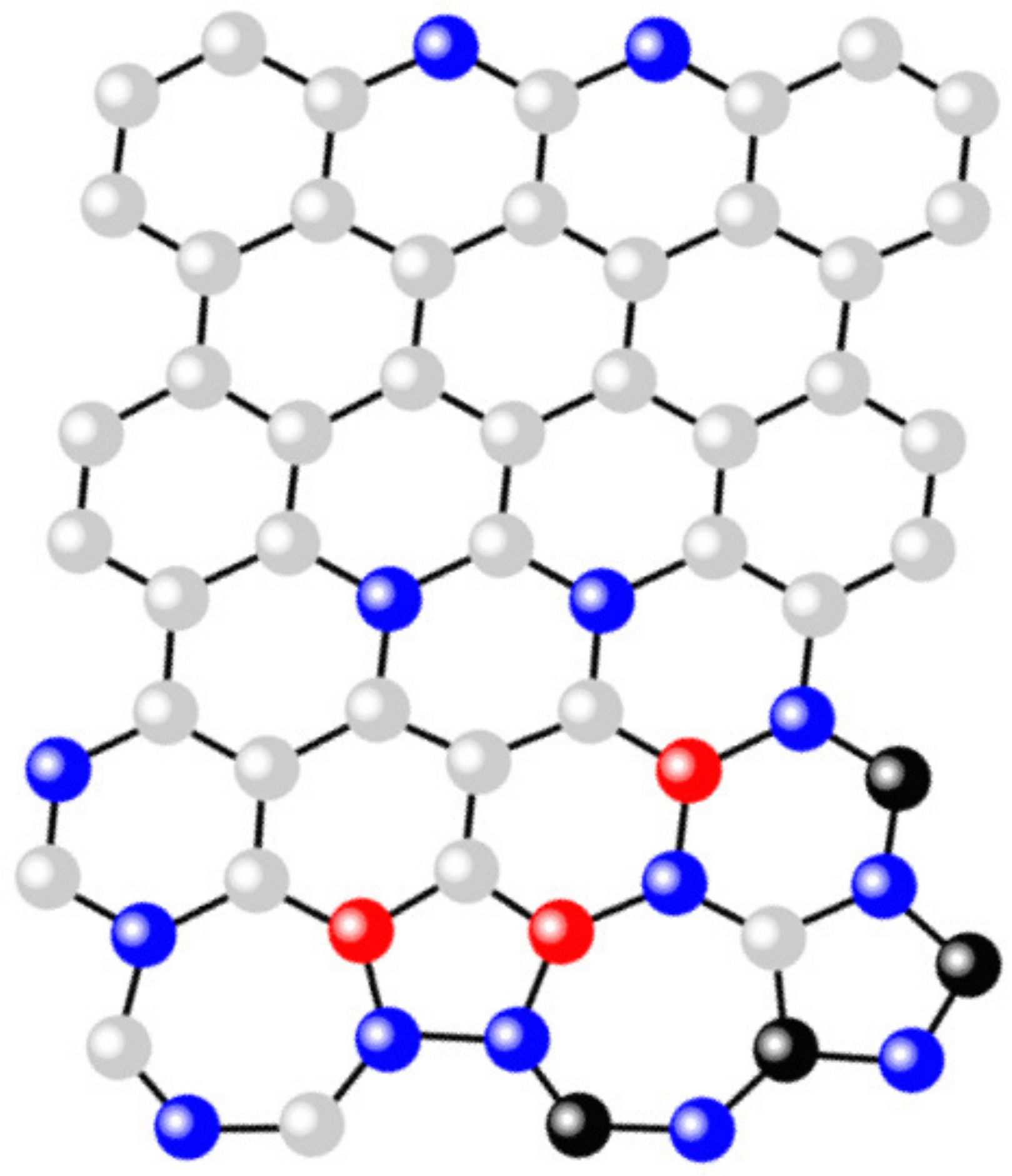}}\subfloat[]{\includegraphics[width=2.2cm]{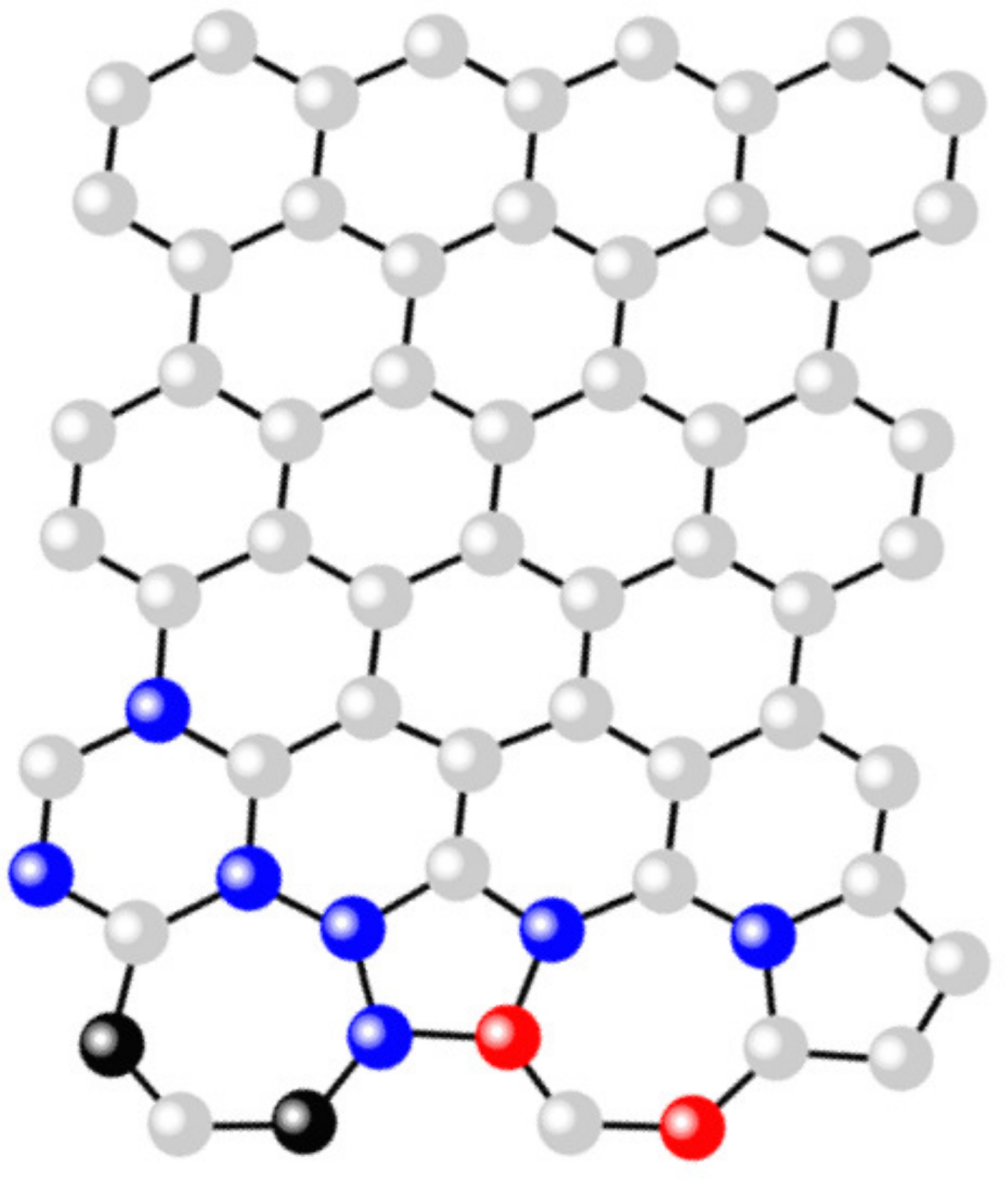}}

\subfloat[]{\includegraphics[width=2.2cm]{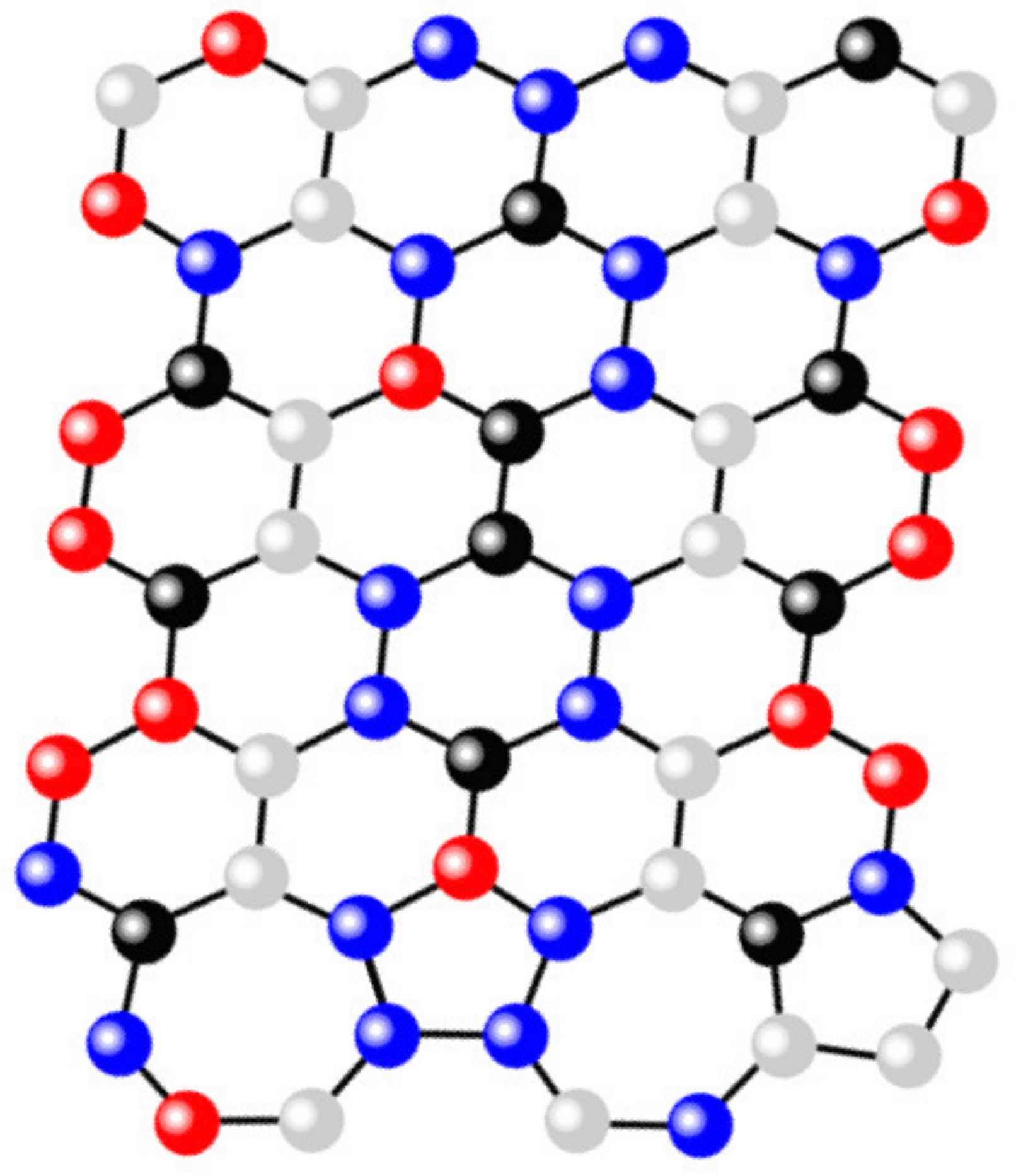}}\subfloat[]{\includegraphics[width=2.2cm]{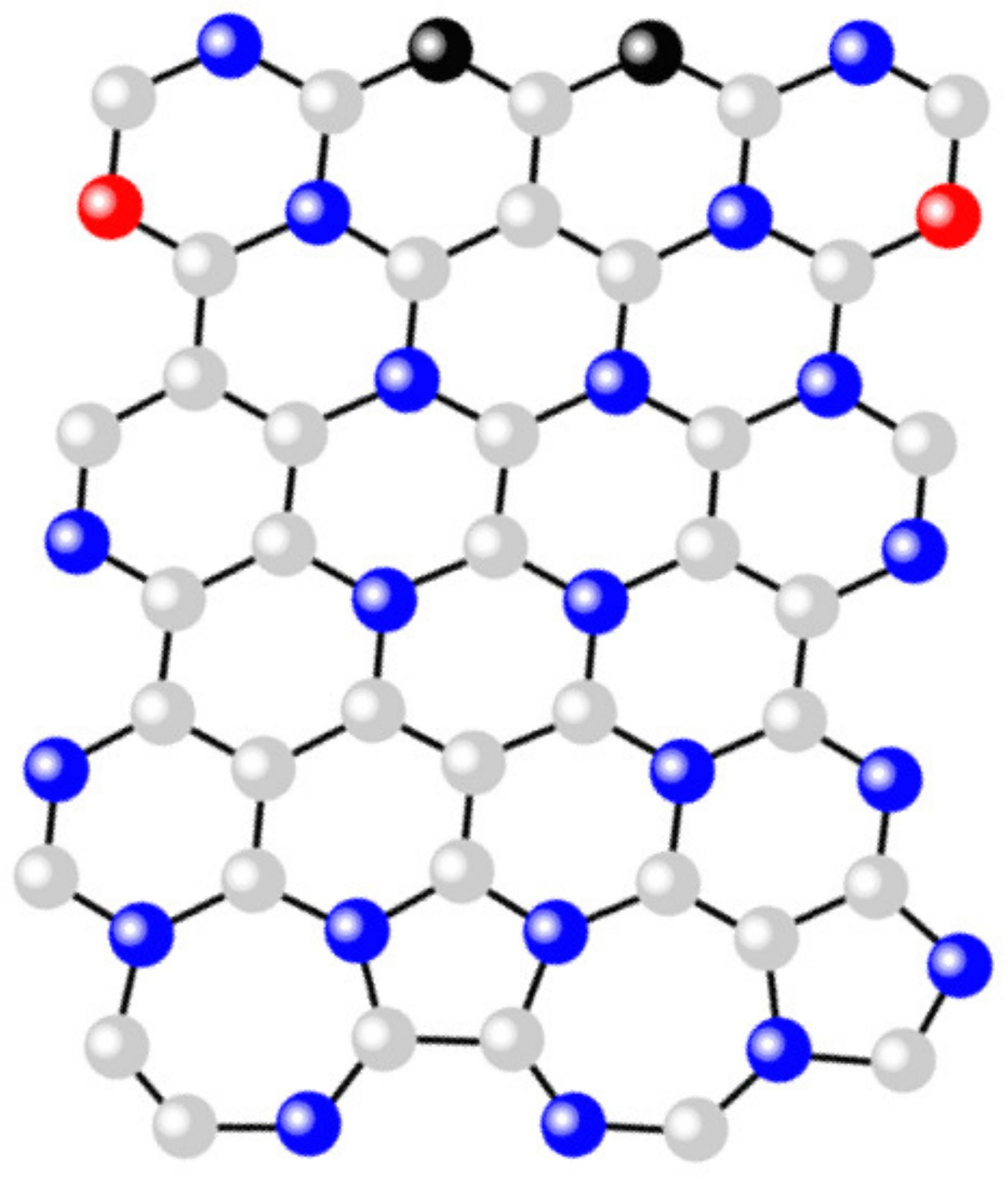}}\subfloat[]{\includegraphics[width=2.2cm]{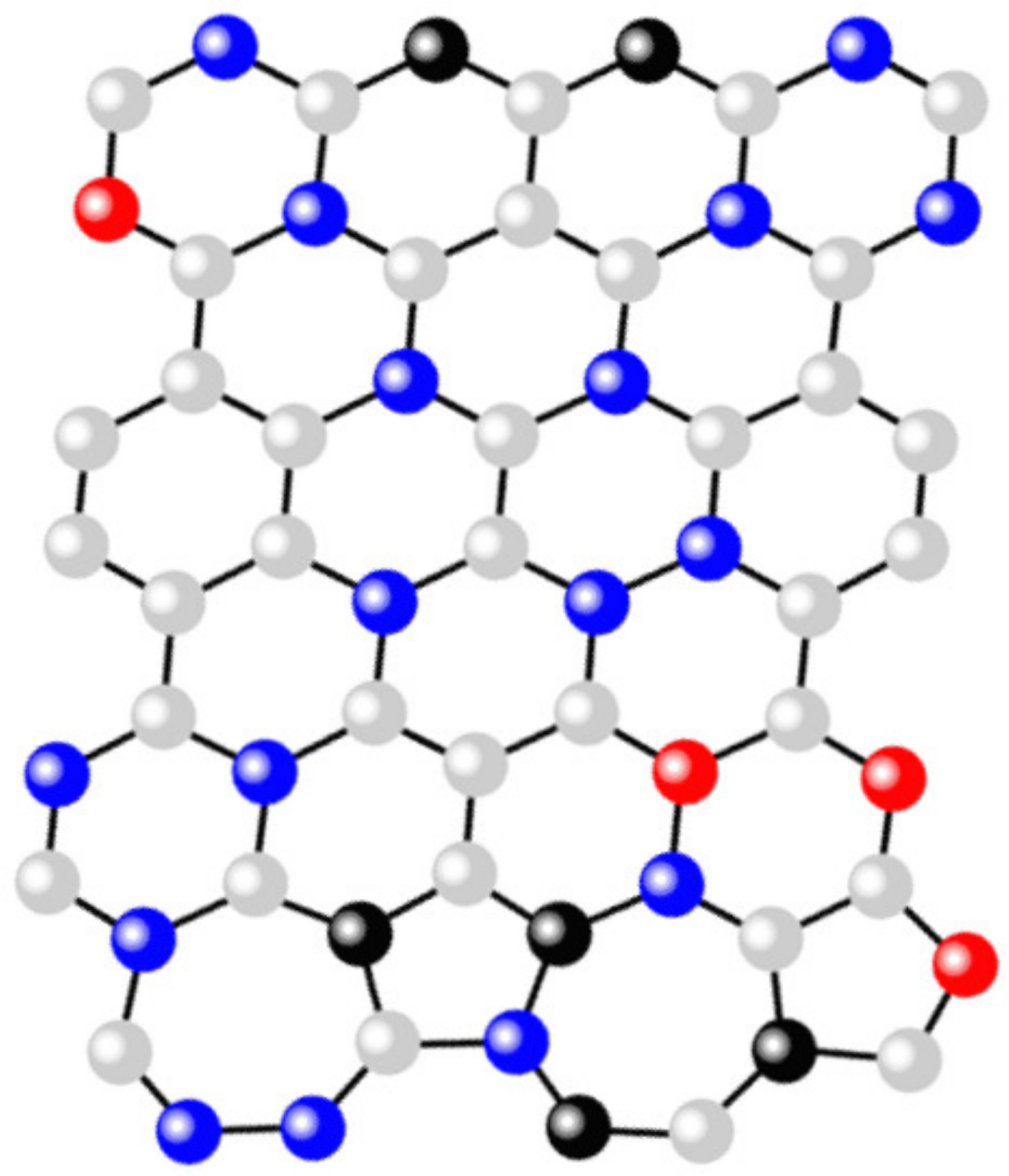}}\subfloat[]{\includegraphics[width=2.2cm]{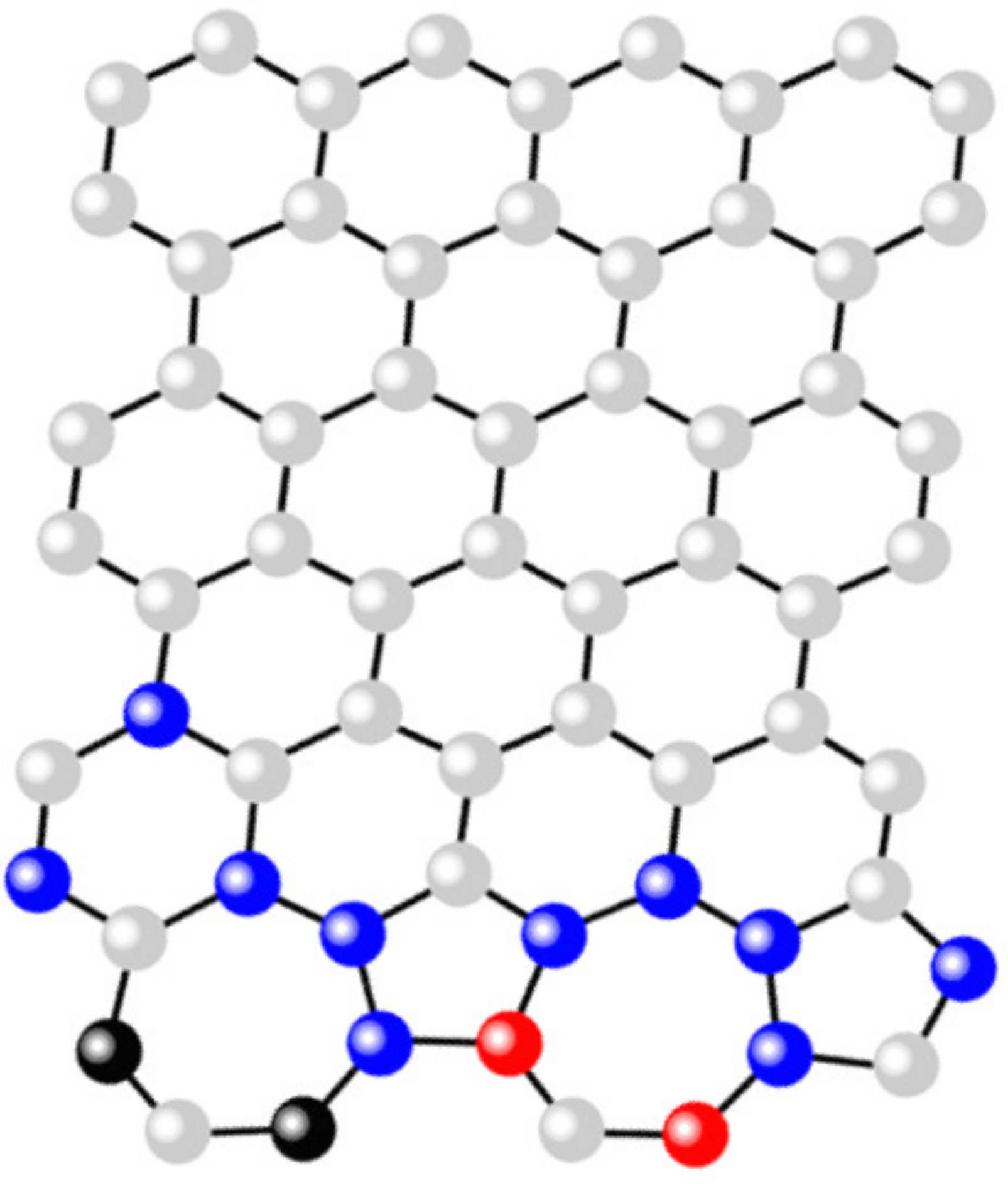}}

\subfloat[]{\includegraphics[width=2.2cm]{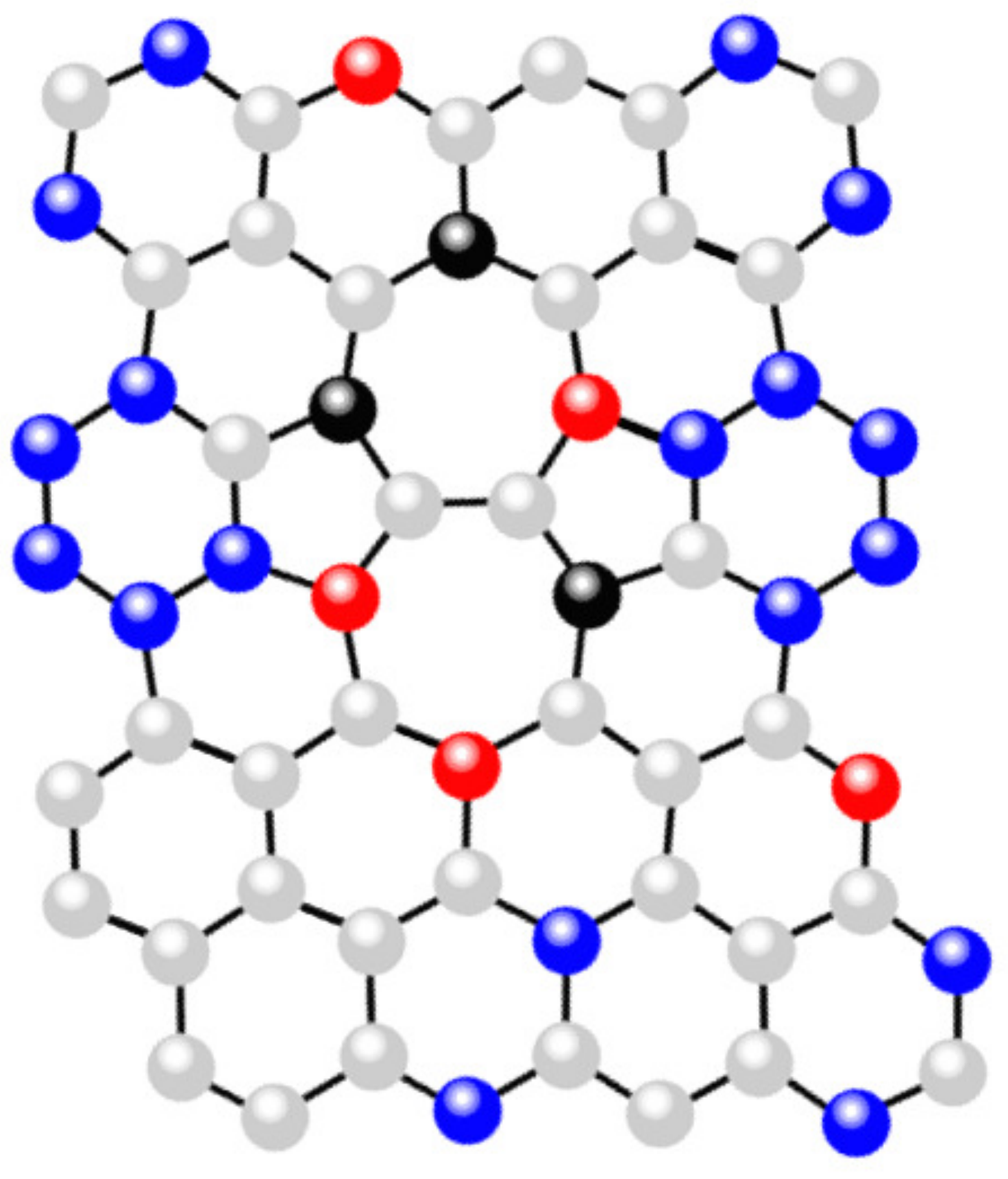}}\subfloat[]{\includegraphics[width=2.2cm]{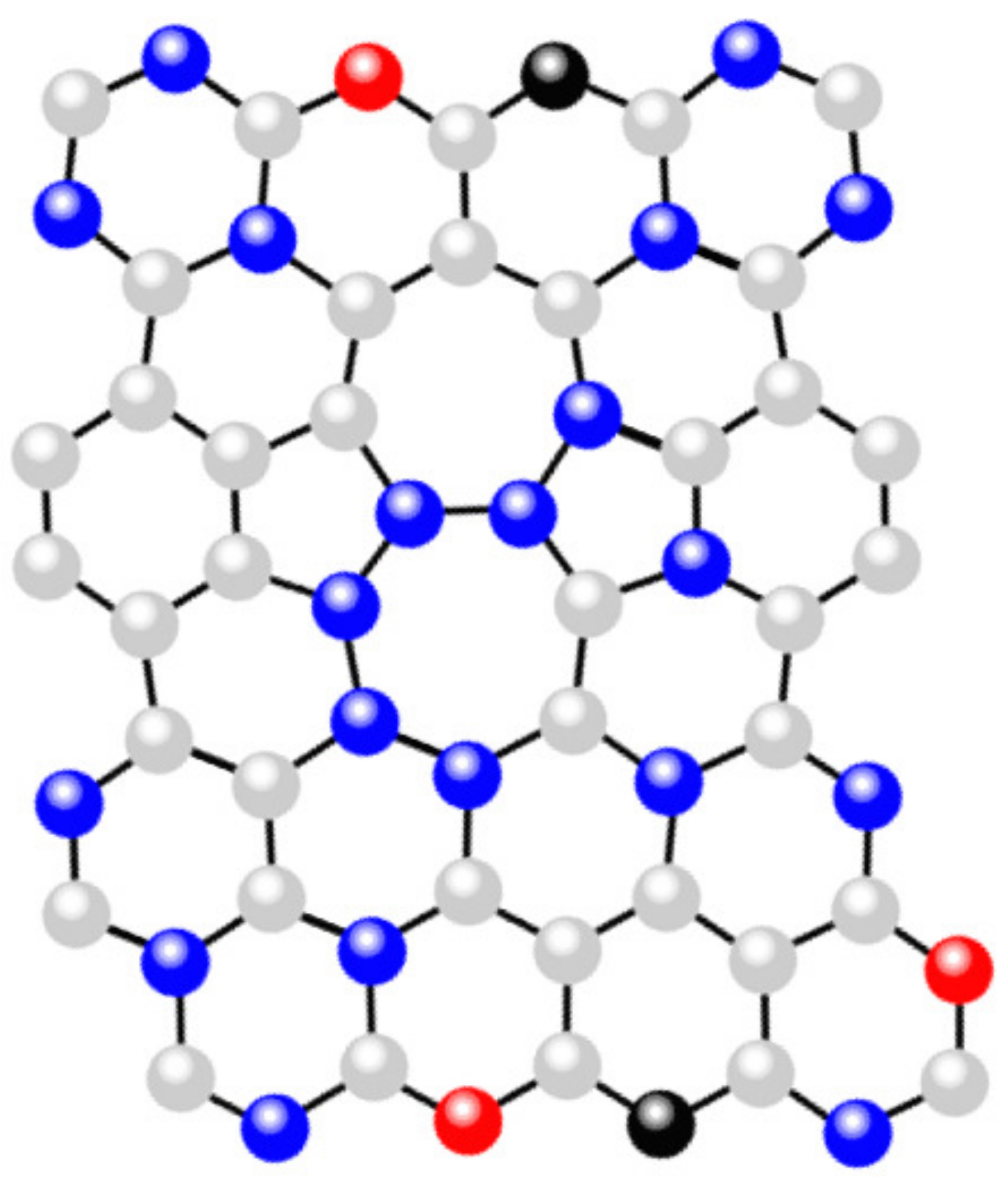}}\subfloat[]{\includegraphics[width=2.2cm]{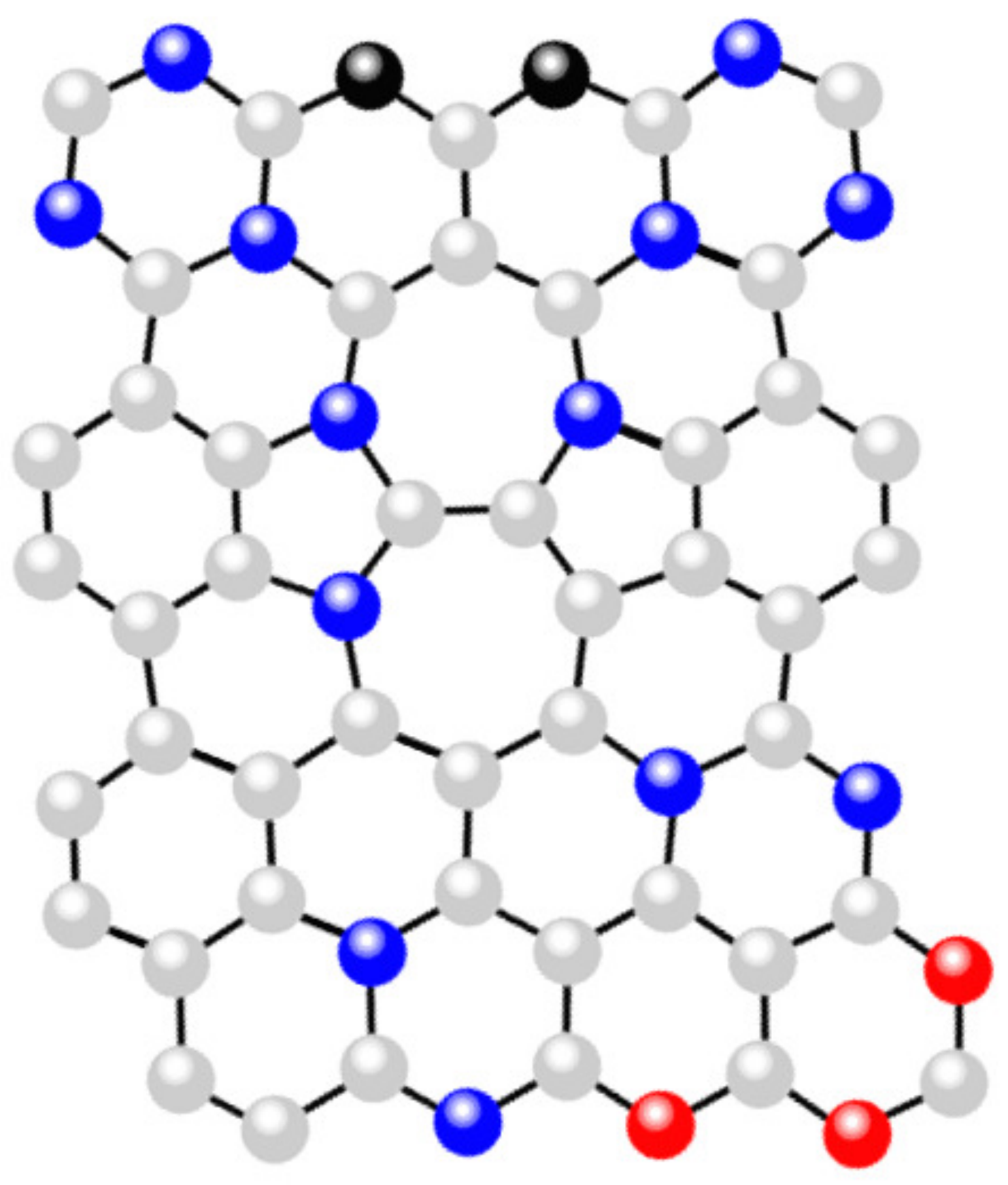}}\subfloat[]{\includegraphics[width=2.2cm]{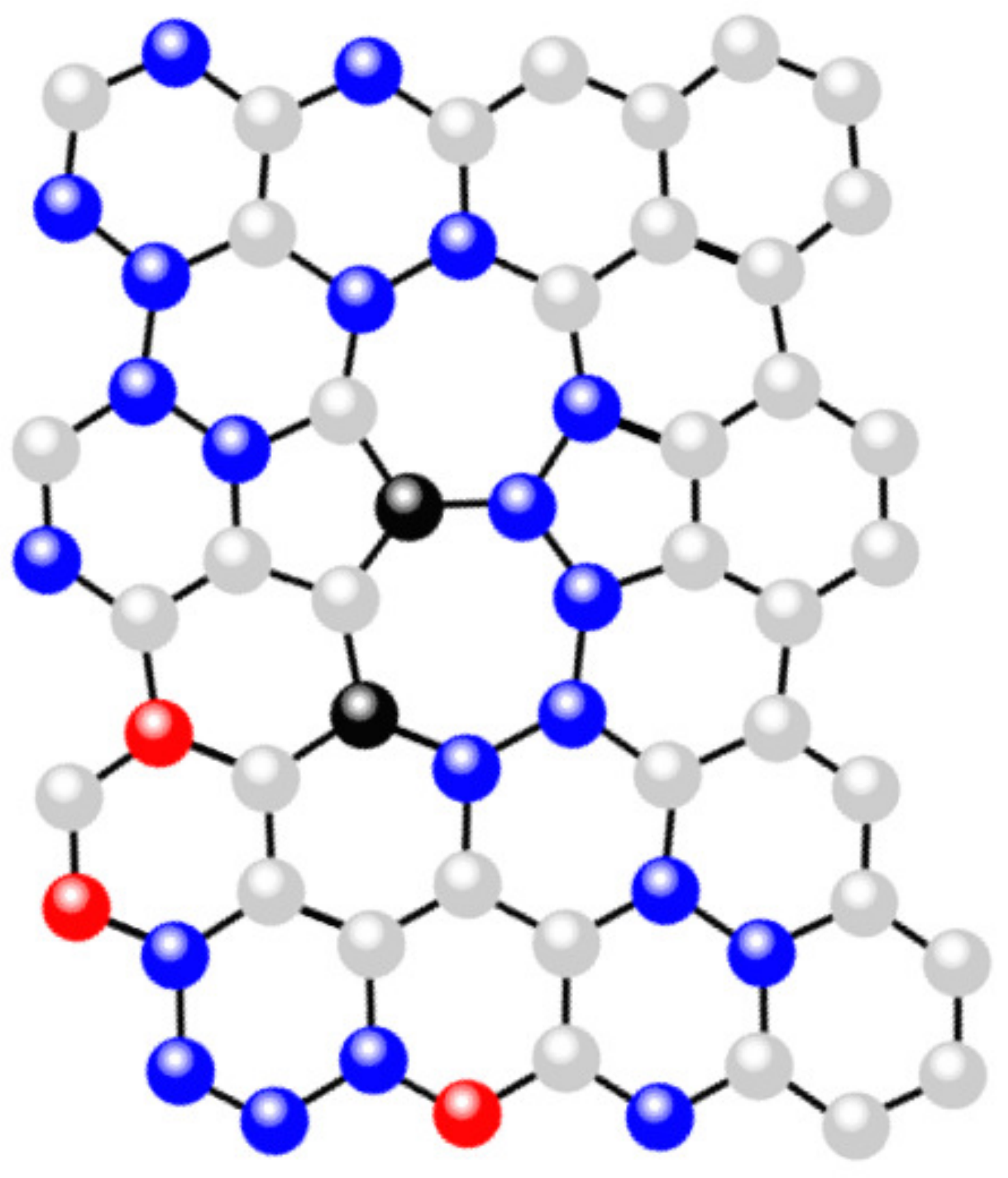}}

\subfloat[]{\includegraphics[width=2.2cm]{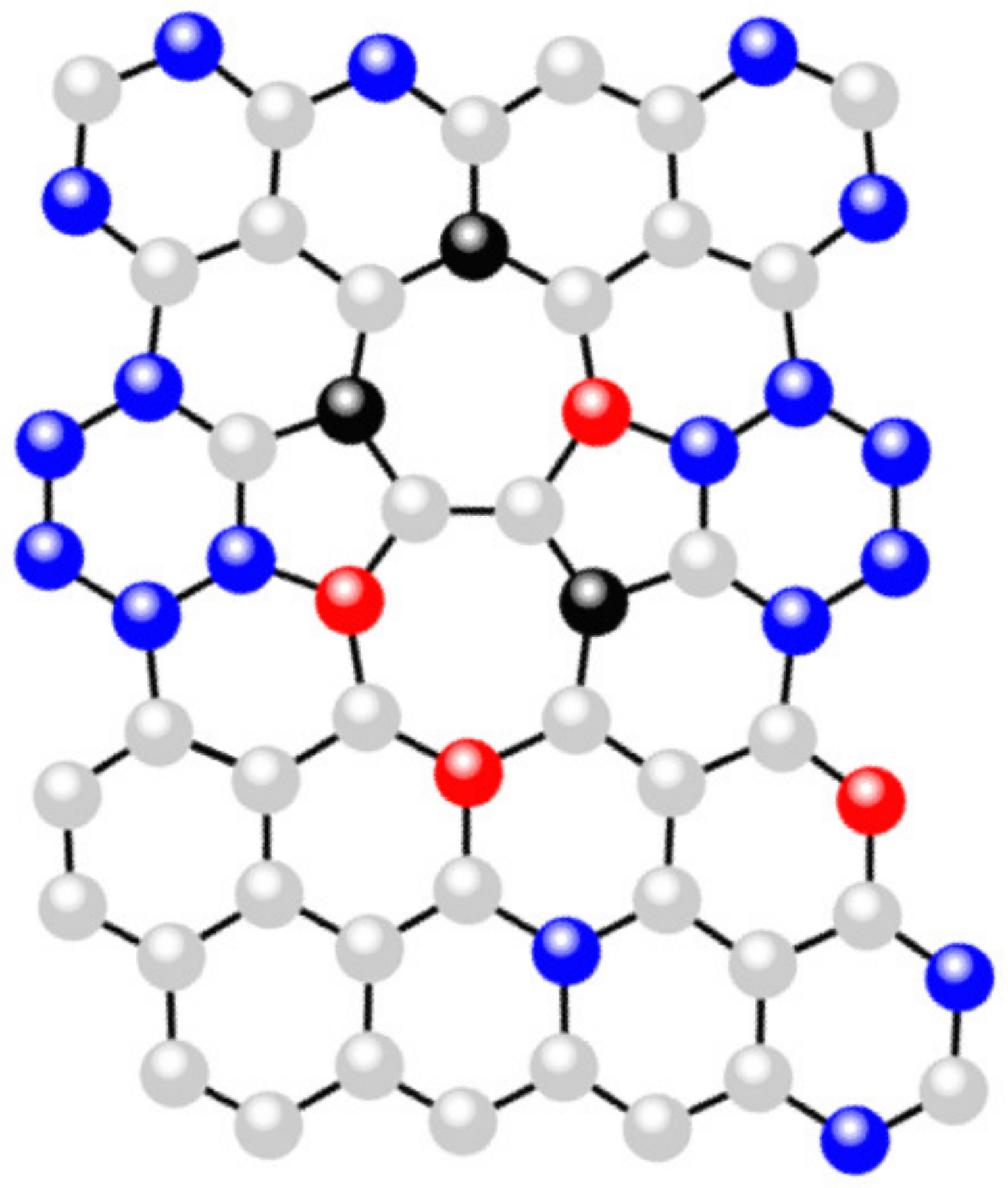}}\subfloat[]{\includegraphics[width=2.2cm]{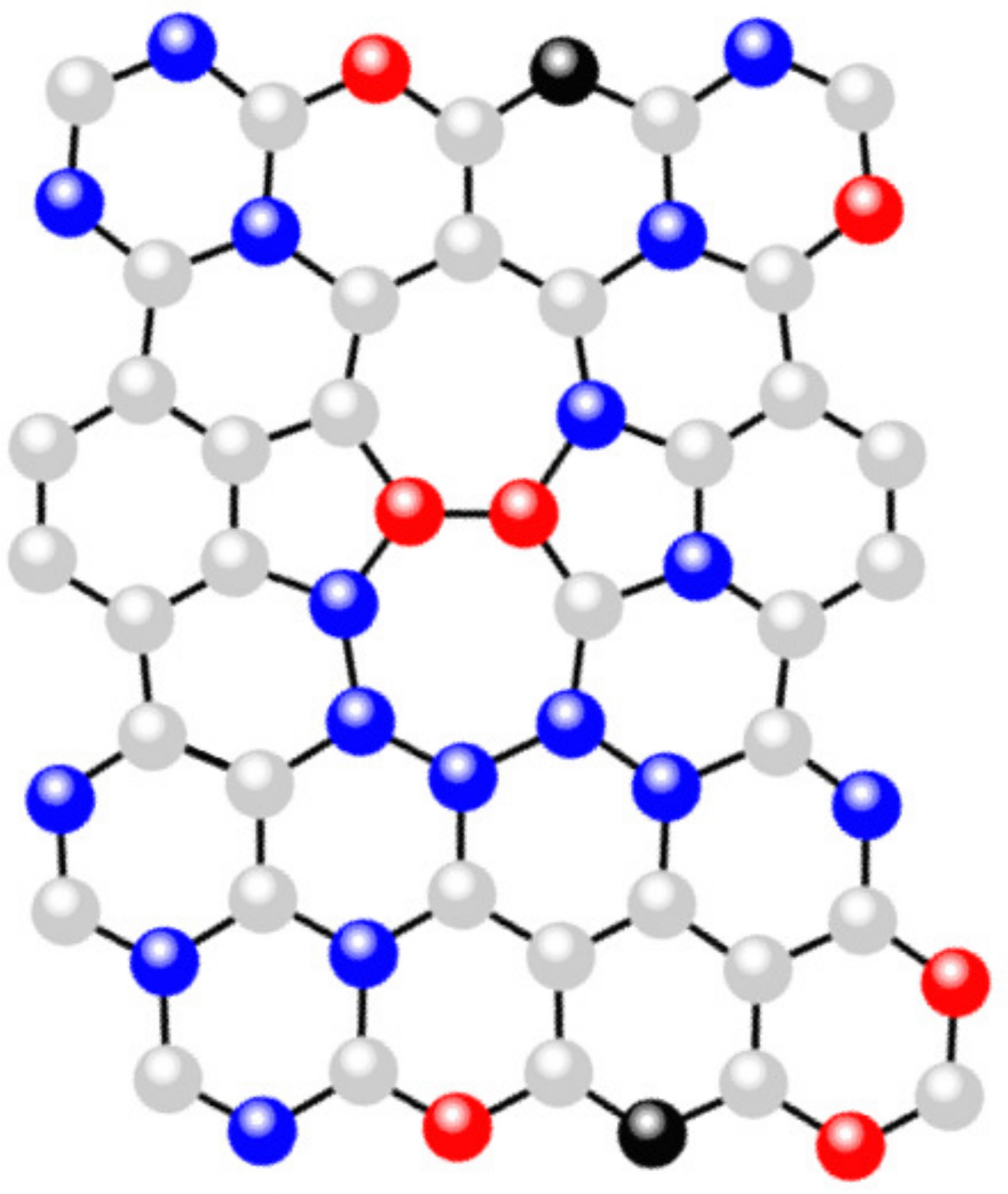}}\subfloat[]{\includegraphics[width=2.2cm]{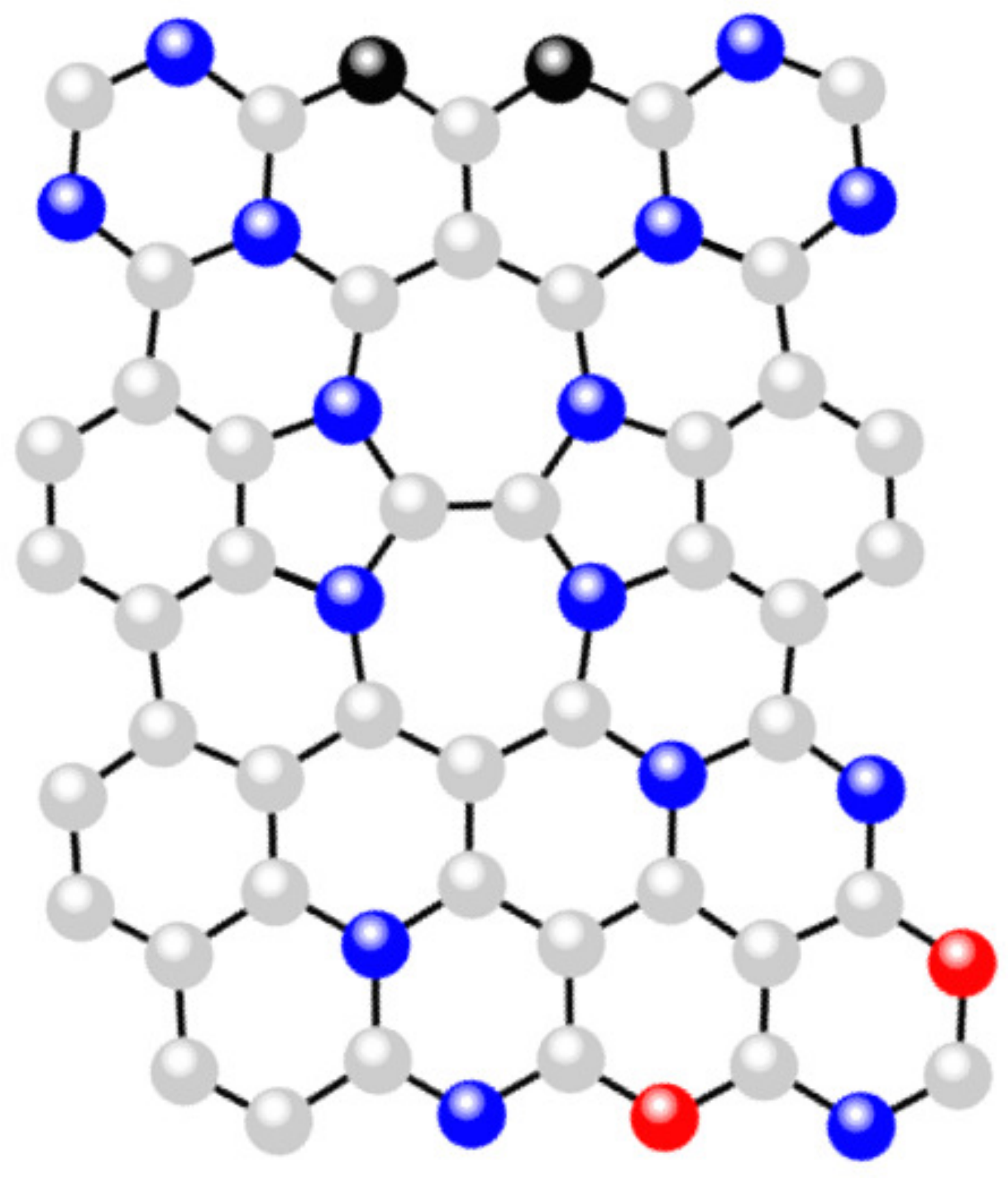}}\subfloat[]{\includegraphics[width=2.2cm]{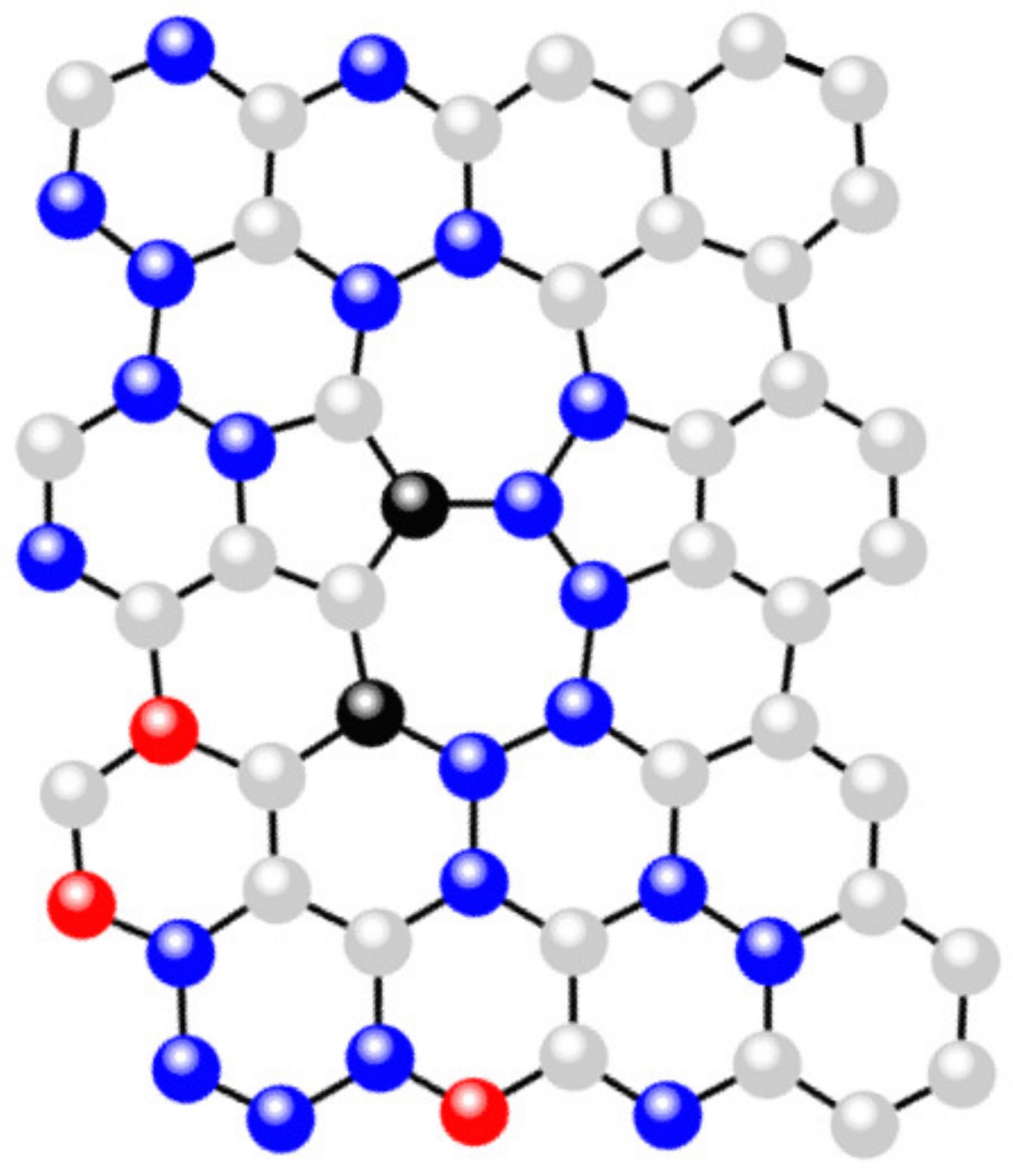}}

\caption{Charge density plots \label{fig:Charge-density-plots} of H-1, H orbitals
of GQD-64 obtained from TB model (a-b) and PPP-HF model (c-d), H-1,
H, L, L+1 orbitals of SW1-GQD-64 and SW2-GQD-64 obtained from TB model
(e-h and m-p) and PPP-HF model (i-l and q-t). The grey, blue, red
and black spheres correspond to $\protect\leq10\%,$ $>10\%-40\%,$$>40\%-70\%$
and >70\% of maximum charge density distribution, respectively.}

\end{figure}

\textcolor{black}{The charge density plots of frontier molecular orbitals
contributing significantly to linear absorption spectra computed from
TB and PPP model for GQD-64, SW1-GQD-64 and SW2-GQD-64 systems are
represented in figs. \ref{fig:Charge-density-plots}(a-t). In case
of GQD-64, the charge density plots of LUMO (L) and L+1 are same as
HOMO (H) and H-1 orbitals, respectively, due to the electron-hole
symmetry, and hence have not been presented. For GQD-64 (figs. \ref{fig:Charge-density-plots}(a-d)),
the charge is concentrated at the zigzag edges of H (or L) orbital,
while for H-1 (or L+1), it is also significant at other atomic sites.
In the presence of SW defects (Figs. \ref{fig:Charge-density-plots}
(e-t)), the electron-hole symmetry is no longer conserved in these
orbital pairs. For SW1-GQD-64 (figs. \ref{fig:Charge-density-plots}(e-l)),
the charge distribution is more conspicuous at the zigzag edge for
H orbital, while it is more at both the reczag and zigzag edges for
L orbital. Further, for H-1 orbital, the charge is quite uniformly
distributed over all atomic sites, which is in contrast to the localized
nature of charge density near the reczag edge for L+1 orbital. In
case of SW2-GQD-64 configuration (figs. \ref{fig:Charge-density-plots}(m-t)),
the charge density is more at the zigzag edge both for H and L orbitals,
while at the location of SW defect, it is high for H, H-1 and L+1
orbitals. This charge concentration at the location of SW defects
suggests that they will play an important role in determining the
electronic band gaps and optical properties of the defective GQDs.}

\subsection{Linear absorption spectrum}

\begin{figure}
\subfloat{\includegraphics[scale=0.26]{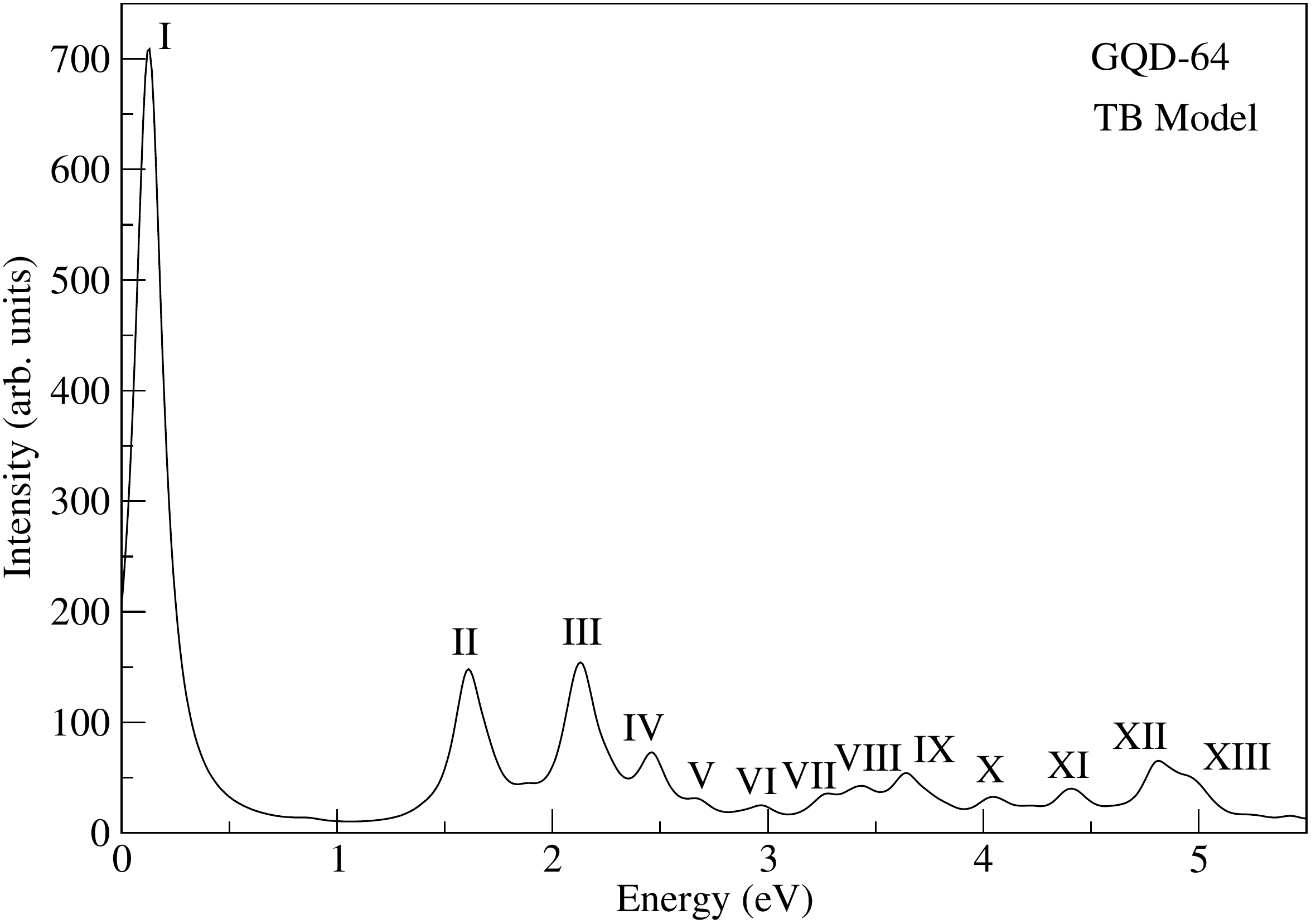}}~~\subfloat{\includegraphics[scale=0.26]{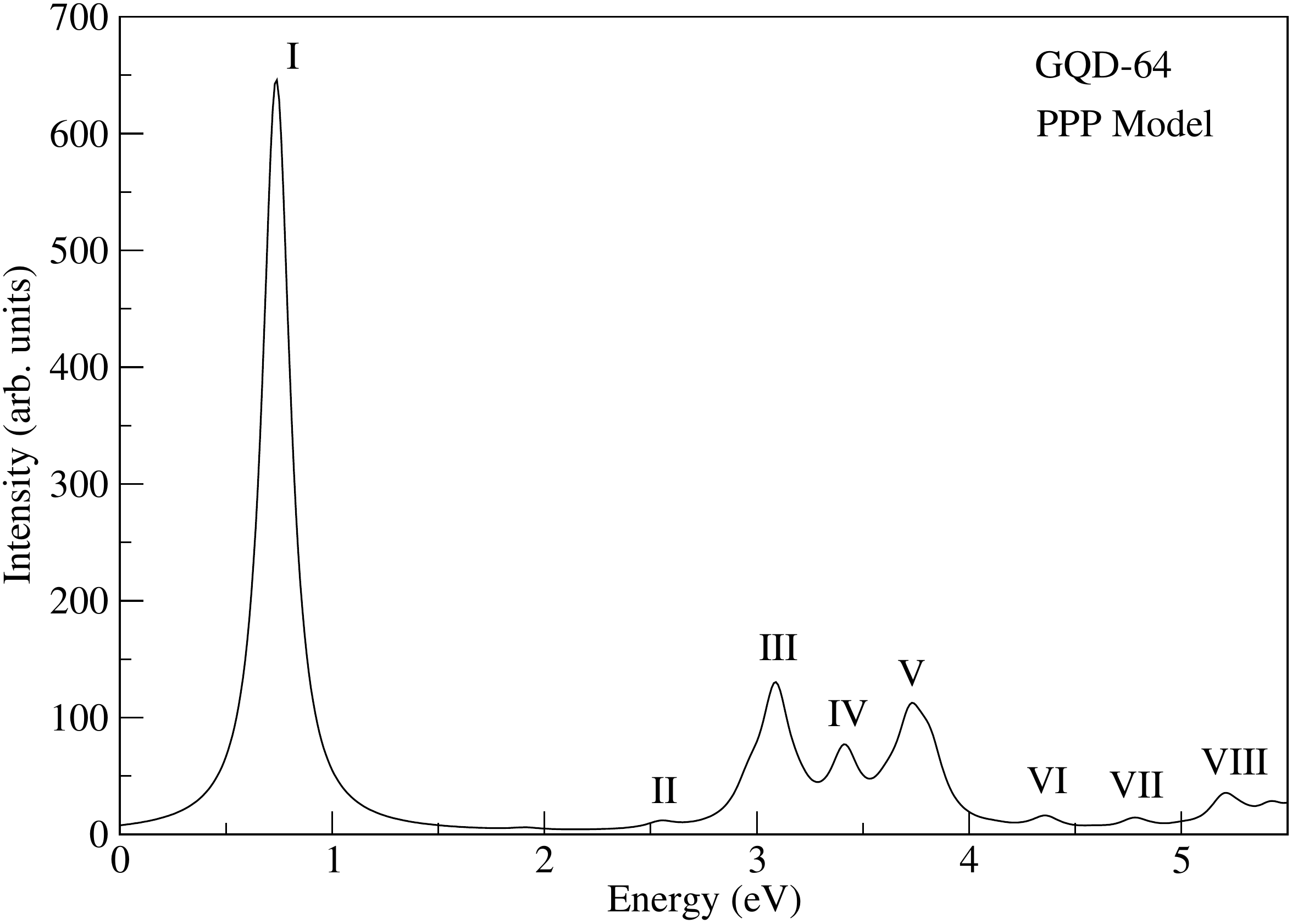}}

\medskip{}

\medskip{}

\subfloat{\includegraphics[scale=0.26]{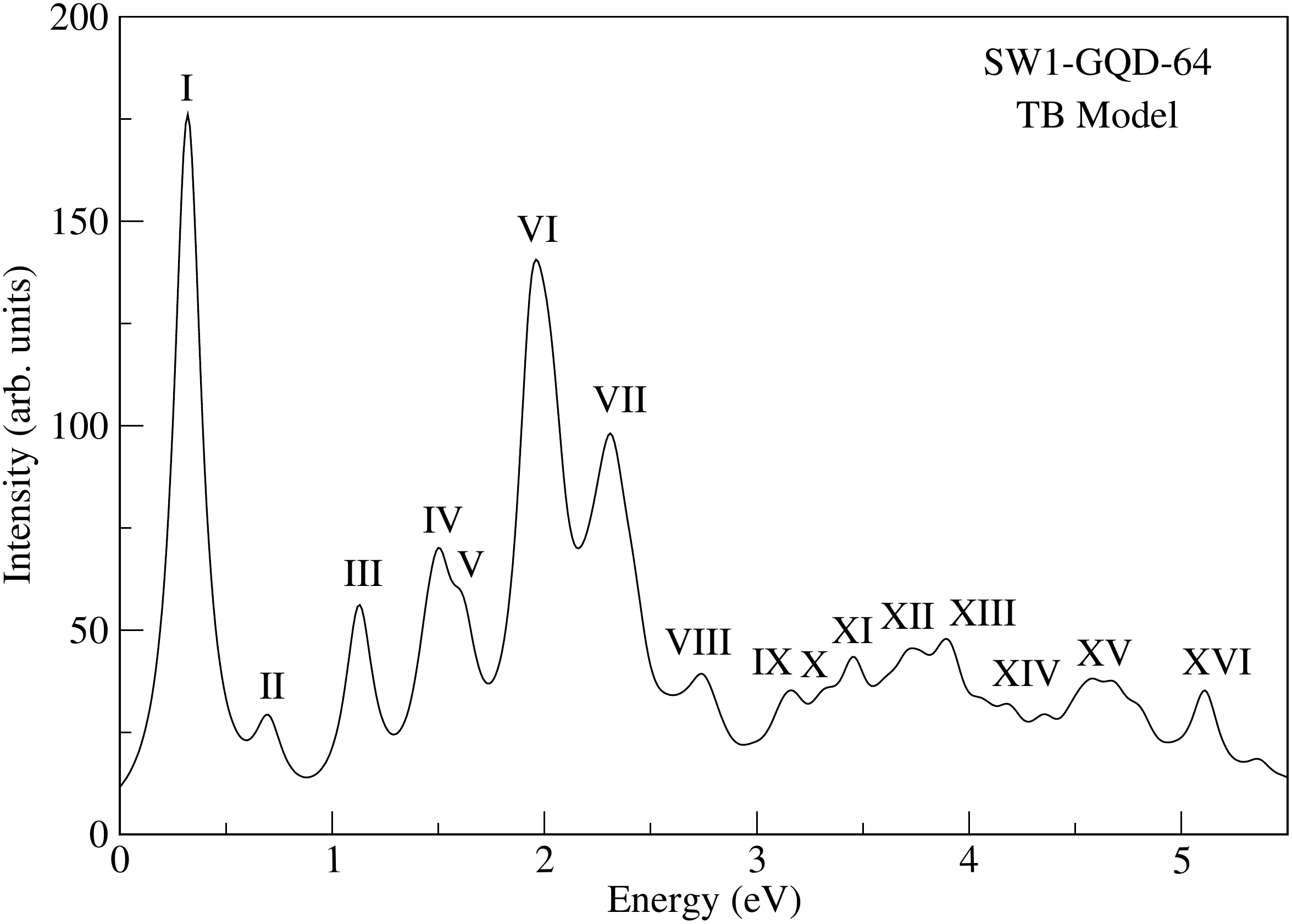}}~~\subfloat{\includegraphics[scale=0.26]{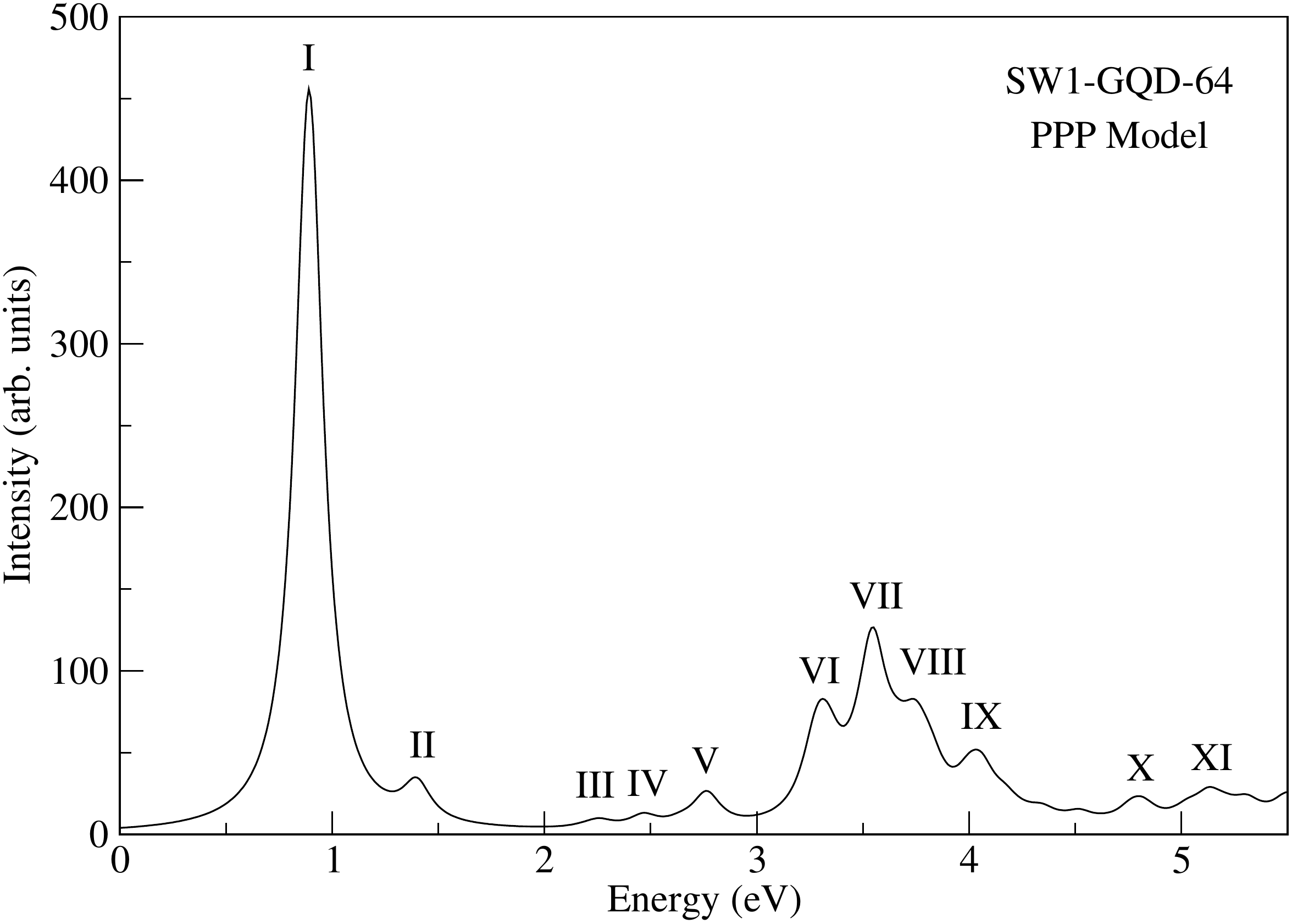}}

\medskip{}

\medskip{}

\subfloat{\includegraphics[scale=0.26]{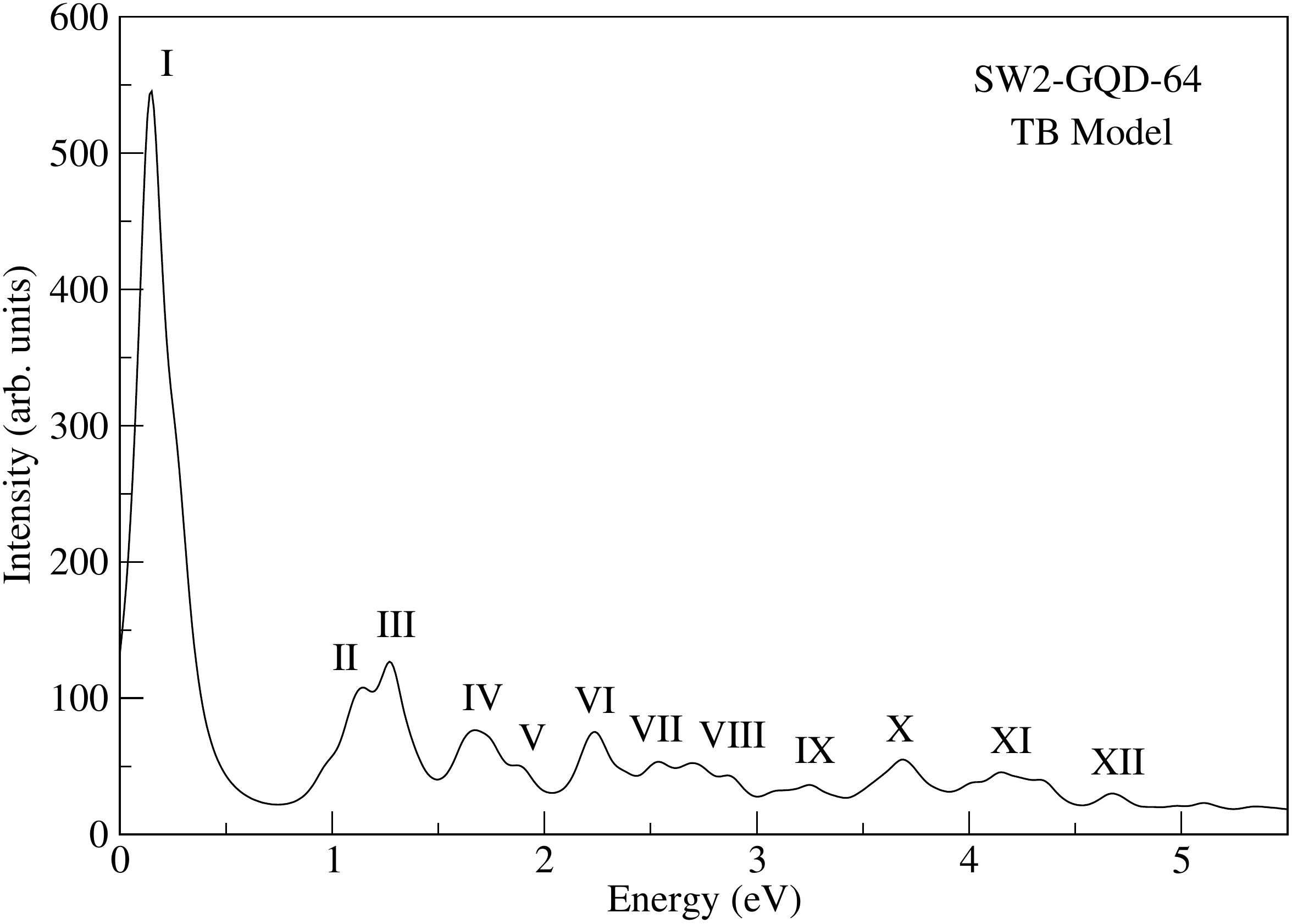}}~\ \subfloat{\includegraphics[scale=0.26]{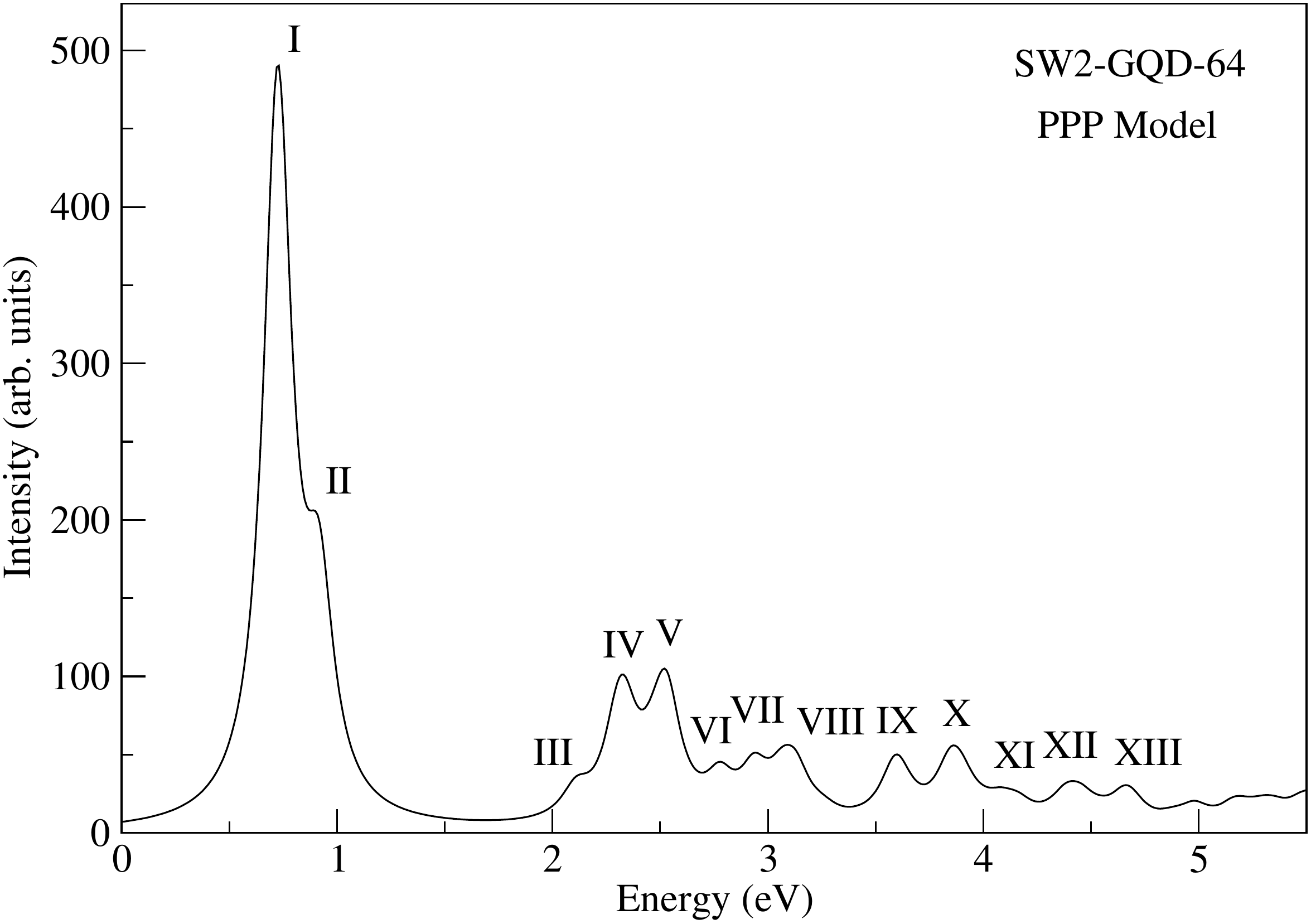}}

\caption{Linear optical absorption spectra of GQD-64, SW1-GQD-64 and SW2-GQD-64
computed using the tight-binding (TB) model and the PPP model at the
HF level. The spectrum has been broadened with a uniform line width
of 0.1 eV.\label{fig:Linear-optical-absorption}}
\end{figure}

\begin{figure}
\subfloat[\label{fig:GQD-64-mrsdci}]{\includegraphics[scale=0.26]{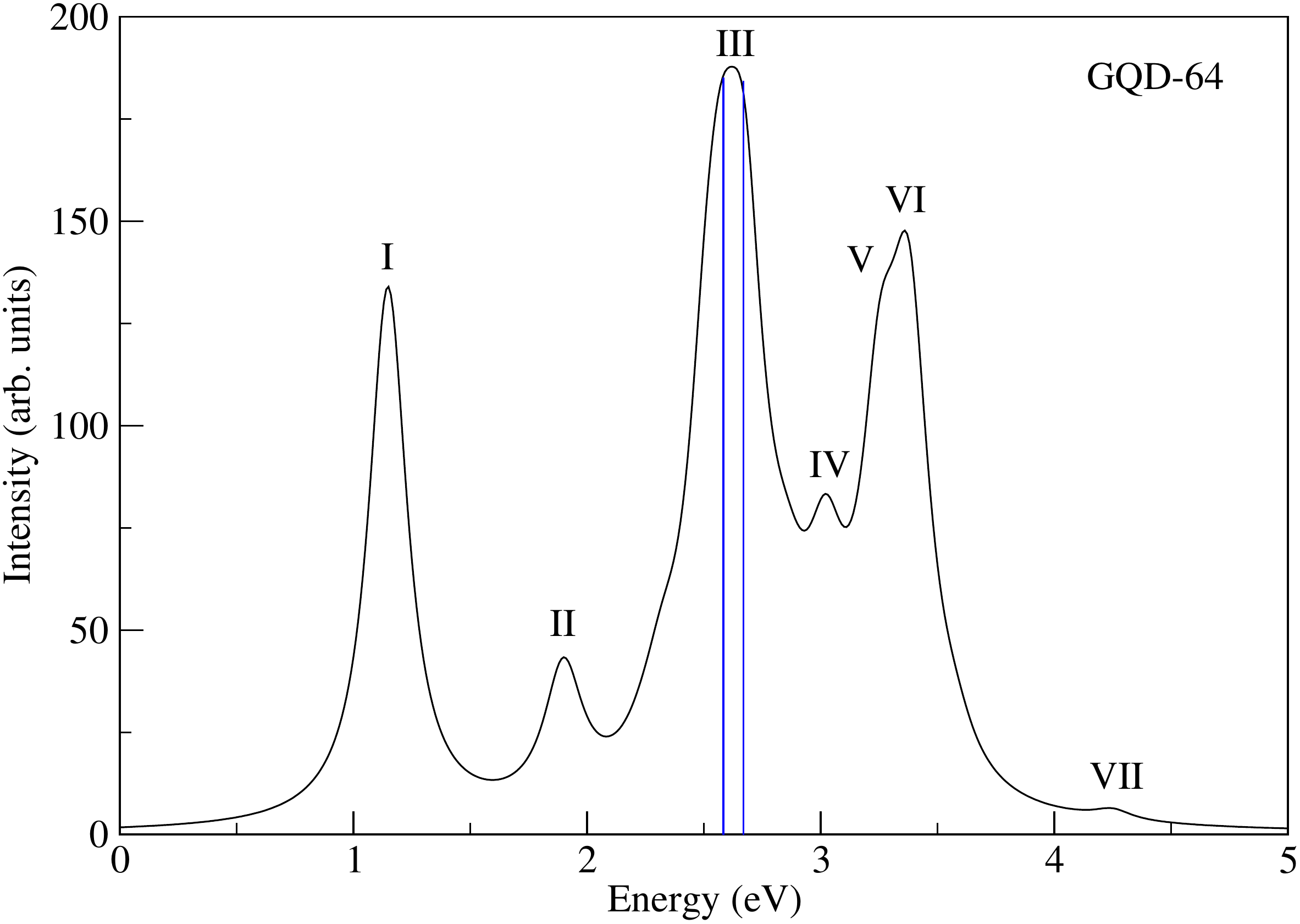}}

\medskip{}

\medskip{}

\subfloat[\label{fig:SW1-GQD-64-mrsdci}]{\includegraphics[scale=0.26]{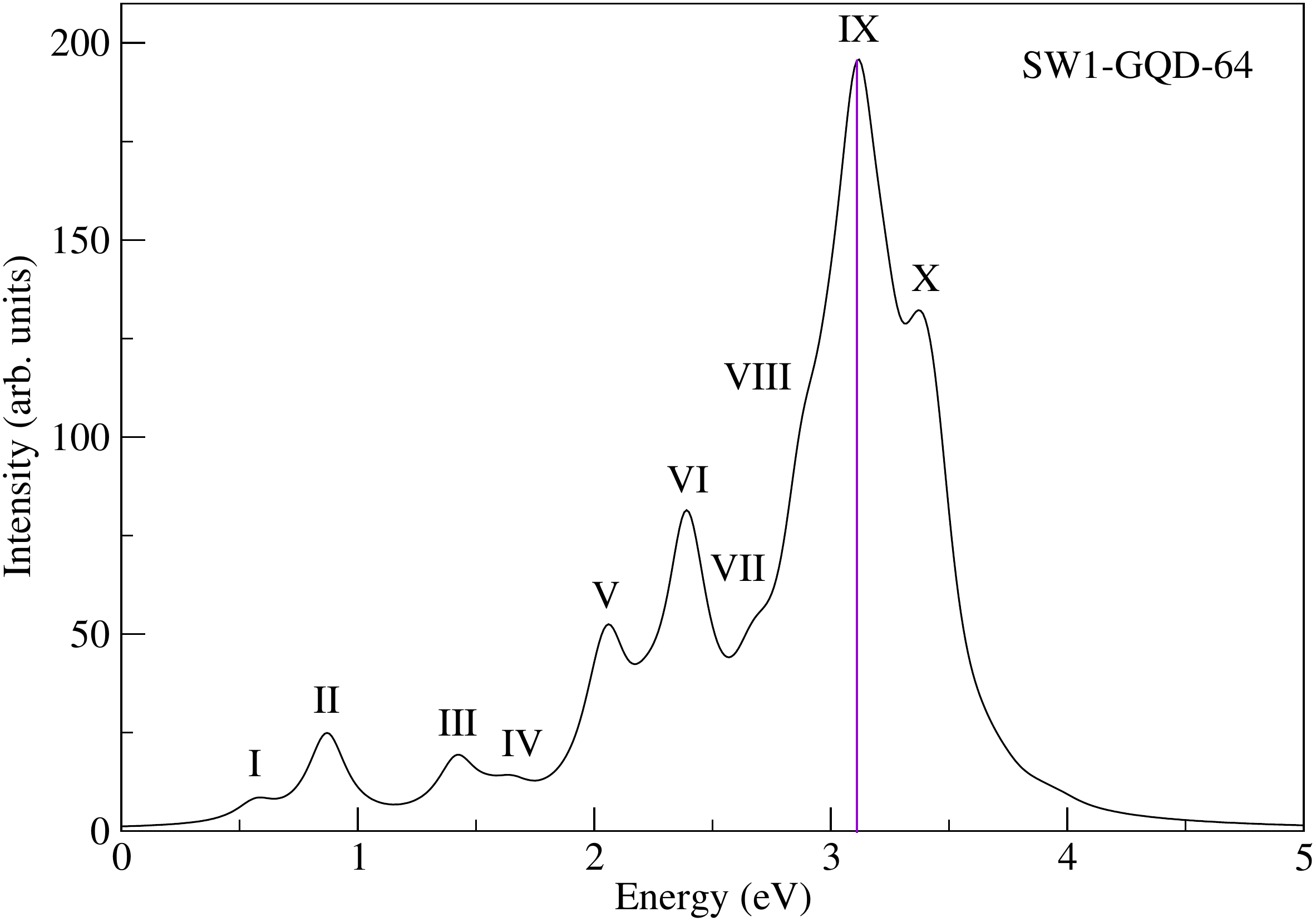}}

\medskip{}

\medskip{}

\subfloat[\label{fig:SW2-GQD-64-mrsdci}]{\includegraphics[scale=0.26]{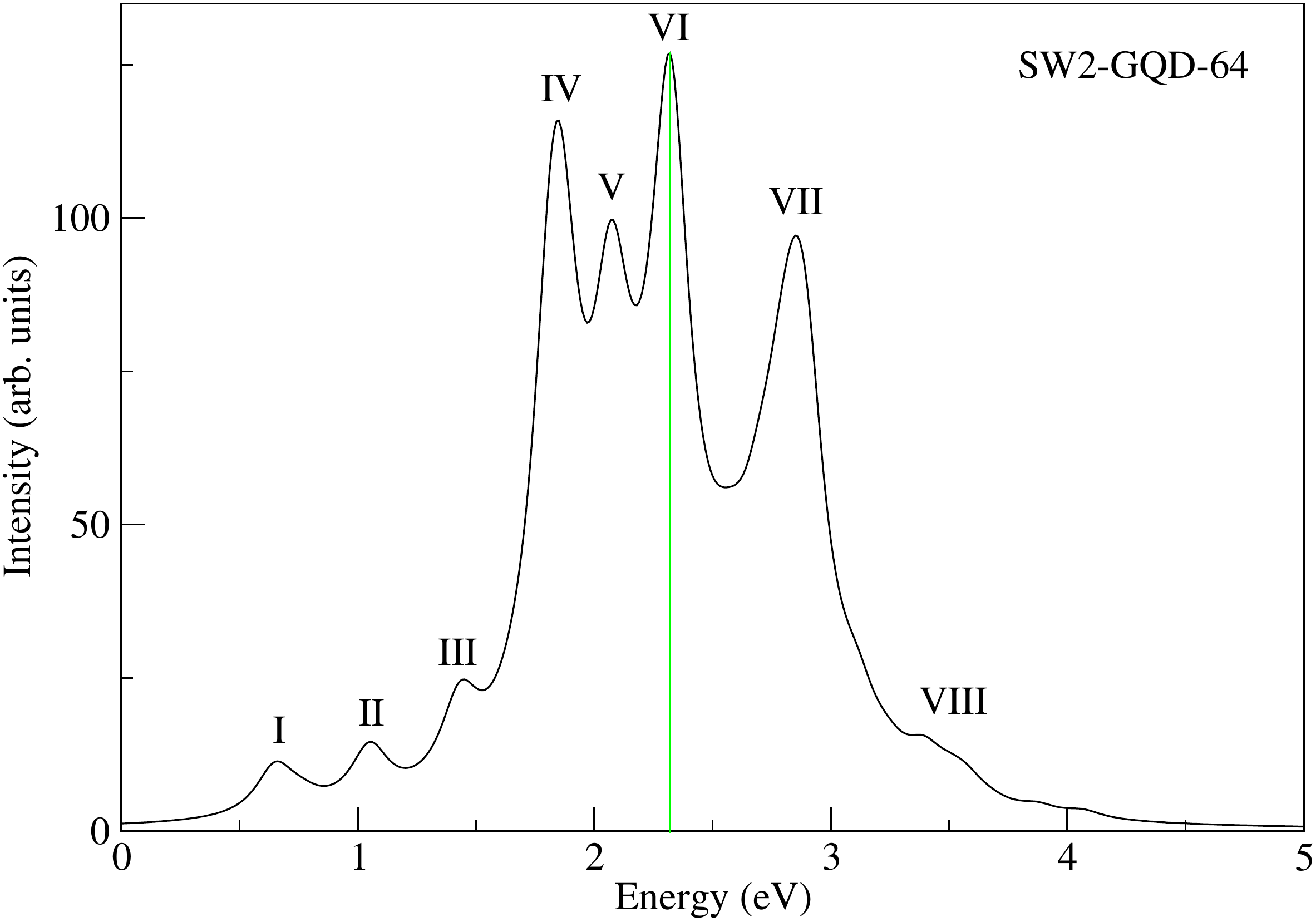}}

\caption{Linear optical absorption spectra of (a) GQD-64, (b) SW1-GQD-64 and
(c) SW2-GQD-64 computed by employing PPP-CI methodology. The spectrum
has been broadened with a uniform line width of 0.1 eV.\textcolor{black}{{}
The blue, violet, and the green lines denote the locations of the
most intense absorption peaks of these systems.}}
\end{figure}

The significance of electron correlation effects in determining the
optical properties of these GQDs are now demonstrated below by critically
analyzing the linear absorption spectra computed using the PPP-CI
methodology (\textcolor{black}{Fig}s. \textcolor{black}{\ref{fig:GQD-64-mrsdci},
\ref{fig:SW1-GQD-64-mrsdci} and \ref{fig:SW2-GQD-64-mrsdci}}) with
those obtained employing the TB and PPP-HF (Fig. \ref{fig:Linear-optical-absorption})
approaches. The reasonably large ($\sim$$10^{6}$) sizes of the CI
matrices considered for these GQDs (\textcolor{black}{Table }\ref{tab:Dimensions-of-the})
indicate that electron correlation effects are well incorporated in
these calculations, thereby signifying the precision of our computations,
since no experimental data on optical absorption spectra is available
till date on these systems. In addition, detailed quantitative information
about the energies, $x$ and $y$ components of the transition dipole
moments, and the dominant configurations which contribute to the many
particle wave-functions of the excited states primarily responsible
for the PPP-CI linear absorption peaks of GQD-64, SW1-GQD-64 and SW2-GQD-64,
are listed in tables \ref{tab:Excited-states-contributing-GQD-64},
\ref{tab:Excited-states-contributing-SW1-GQD-64} and \ref{tab:Excited-states-contributing-SW2-GQD-64},
respectively. 

\begin{table}
\caption{Dimensions of the CI matrices\label{tab:Dimensions-of-the} for computing
the linear absorption spectra of GQD-64, SW1-GQD-64 and SW2-GQD-64.}

\begin{tabular}{|c|c|}
\hline 
System & Dimension of CI Matrix\tabularnewline
\hline 
\hline 
GQD-64 & 6118740\tabularnewline
\hline 
SW1-GQD-64 & 7660787\tabularnewline
\hline 
SW2-GQD-64 & 8135639\tabularnewline
\hline 
\end{tabular}
\end{table}

\begin{table}
\caption{\textcolor{black}{Excited states contributing to the linear absorption
spectrum of GQD-64\label{tab:Excited-states-contributing-GQD-64},
computed using the MRSDCI methodology. Here, $c.c.$ implies the charge-conjugated
configuration.}}

\begin{tabular}{|c|c|c|c|c|}
\hline 
Peak  & E (eV)  & Transition  & Transition  & Dominant Configurations\tabularnewline
 &  & Dipole {\small{}(Å) x-component} & Dipole {\small{}(Å) y-component} & \tabularnewline
\hline 
$I$ & $1.15$ & $0.70$ & $3.31$ & $|H\rightarrow L\rangle$$(0.8333)$\tabularnewline
 &  &  &  & $|H-1\rightarrow L+1\rangle$$(0.1650)$\tabularnewline
\hline 
$II$ & $1.90$ & $1.30$ & $0.27$ & $|H\rightarrow L;H\rightarrow L+1\rangle$$+c.c.(0.5665)$ \tabularnewline
 &  &  &  & $|H-3\rightarrow L\rangle$$+c.c.(0.1641)$ \tabularnewline
\hline 
$III$  & $2.58$ & $0.19$ & $1.70$ & $|H-1\rightarrow L+1\rangle$$(0.4903)$\tabularnewline
 &  &  &  & $|H-3\rightarrow L\rangle$$+c.c.(0.3696)$\tabularnewline
 & $2.67$ & $0.85$ & $1.80$ & $|H-1\rightarrow L+1\rangle$$(0.5933)$\tabularnewline
 &  &  &  & $|H-3\rightarrow L\rangle$$+c.c.(0.3857)$\tabularnewline
\hline 
$IV$ & $3.02$ & $1.17$ & $0.04$ & $|H-5\rightarrow L\rangle$$+c.c.(0.3681)$\tabularnewline
 &  &  &  & $|H-6\rightarrow L\rangle$$-c.c.(0.3513)$\tabularnewline
\hline 
$V$  & $3.25$ & $0.95$ & $1.17$ & $|H-5\rightarrow L\rangle$$+c.c.(0.4083)$\tabularnewline
 &  &  &  & $|H\rightarrow L+6\rangle$$-c.c.(0.3326)$\tabularnewline
\hline 
$VI$  & $3.38$ & $1.69$ & $0.06$ & $|H\rightarrow L+7\rangle$$-c.c.(0.5507)$\tabularnewline
 &  &  &  & $|H\rightarrow L+10\rangle$$+c.c.(0.1321)$\tabularnewline
\hline 
$VII$ & $4.25$ & $0.22$ & $0.11$ & $|H\rightarrow L+11\rangle$$-c.c.(0.4248)$\tabularnewline
 &  &  &  & $|H\rightarrow L;H\rightarrow L+4\rangle$$-c.c.(0.4001)$\tabularnewline
\hline 
\end{tabular}
\end{table}

\begin{table}
\caption{\textcolor{black}{Excited states contributing to the linear absorption
spectrum of SW1-GQD-64\label{tab:Excited-states-contributing-SW1-GQD-64},
computed employing the MRSDCI methodology.}}

\begin{tabular}{|c|c|c|c|c|}
\hline 
Peak  & E (eV)  & Transition  & Transition  & Dominant Configurations\tabularnewline
 &  & Dipole {\small{}(Å) x-com} & Dipole {\small{}(Å) y-com} & \tabularnewline
\hline 
$I$ & $0.57$ & $0.66$ & $0.65$ & $|H\rightarrow L+1\rangle$$(0.7152)$\tabularnewline
 &  &  &  & $|H\rightarrow L;H\rightarrow L+1\rangle$$(0.4383)$\tabularnewline
\hline 
$II$ & $0.87$ & $0.58$ & $1.50$ & $|H\rightarrow L\rangle$$(0.6061)$\tabularnewline
 &  &  &  & $|H\rightarrow L;H\rightarrow L\rangle$$(0.3784)$\tabularnewline
\hline 
$III$  & $1.42$ & $0.22$ & $0.96$ & $|H\rightarrow L\rangle$$(0.5216)$\tabularnewline
 &  &  &  & $|H\rightarrow L;H\rightarrow L\rangle$$(0.5072)$\tabularnewline
\hline 
$IV$ & $1.65$ & $0.11$ & $0.45$ & $|H-1\rightarrow L\rangle$$(0.5760)$\tabularnewline
 &  &  &  & $|H\rightarrow L+2\rangle$$(0.3075)$\tabularnewline
\hline 
$V$  & $2.06$ & $1.11$ & $0.60$ & $|H\rightarrow L;H\rightarrow L+1\rangle$$(0.4645)$\tabularnewline
 &  &  &  & $|H\rightarrow L;H-1\rightarrow L\rangle$$(0.4073)$\tabularnewline
\hline 
$VI$  & $2.39$ & $1.54$ & $0.24$ & $|H\rightarrow L;H-1\rightarrow L\rangle$$(0.4360)$\tabularnewline
 &  &  &  & $|H\rightarrow L;H\rightarrow L+2\rangle$$(0.4256)$\tabularnewline
\hline 
$VII$ & $2.66$ & $0.16$ & $0.69$ & $|H\rightarrow L+6\rangle$$(0.5775)$\tabularnewline
 &  &  &  & $|H\rightarrow L+5\rangle$$(0.3303)$\tabularnewline
\hline 
$VIII$  & $2.91$ & $0.43$ & $0.30$ & $|H-4\rightarrow L\rangle$$(0.2895)$\tabularnewline
 &  &  &  & $|H\rightarrow L+6\rangle$$(0.2593)$\tabularnewline
 &  &  &  & $|H\rightarrow L+4\rangle$$(0.2554)$\tabularnewline
 &  &  &  & $|H\rightarrow L;H-1\rightarrow L+1\rangle$$(0.2535)$\tabularnewline
\hline 
$IX$  & $3.11$ & $1.06$ & $1.34$ & $|H-1\rightarrow L+2\rangle$$(0.5703)$\tabularnewline
 &  &  &  & $|H-5\rightarrow L\rangle$$(0.4346)$\tabularnewline
\hline 
$X$ & $3.38$ & $0.66$ & $0.48$ & $|H\rightarrow L+11\rangle$$(0.4879)$\tabularnewline
 &  &  &  & $|H-7\rightarrow L\rangle$$(0.3161)$\tabularnewline
\hline 
\end{tabular}
\end{table}

\begin{table}
\caption{\textcolor{black}{Excited states contributing to the linear absorption
spectrum of SW2-GQD-64\label{tab:Excited-states-contributing-SW2-GQD-64},
computed employing the MRSDCI methodology.}}

\begin{tabular}{|c|c|c|c|c|}
\hline 
Peak  & E (eV)  & Transition  & Transition  & Dominant Configurations\tabularnewline
 &  & Dipole {\small{}(Å) x-com} & Dipole {\small{}(Å) y-com} & \tabularnewline
\hline 
$I$ & $0.65$ & $0.97$ & $0.51$ & $|H-1\rightarrow L\rangle$$(0.7675)$\tabularnewline
 &  &  &  & $|H\rightarrow L\rangle$$(0.2601)$\tabularnewline
\hline 
$II$ & $1.05$ & $0.73$ & $0.61$ & $|H\rightarrow L\rangle$$(0.6223)$\tabularnewline
 &  &  &  & $|H-1\rightarrow L\rangle$$(0.2684)$\tabularnewline
\hline 
$III$  & $1.44$ & $0.003$ & $0.19$ & $|H-1\rightarrow L;H-1\rightarrow L\rangle$$(0.5757)$\tabularnewline
 &  &  &  & $|H\rightarrow L;H\rightarrow L\rangle$$(0.3828)$\tabularnewline
\hline 
$IV$  & $1.86$ & $0.92$ & $1.41$ & $|H-1\rightarrow L+2\rangle$$(0.3857)$\tabularnewline
 &  &  &  & $|H-1\rightarrow L+1;H\rightarrow L\rangle$$(0.3640)$\tabularnewline
 &  &  &  & $|H-1\rightarrow L+1\rangle$$(0.3540)$\tabularnewline
\hline 
$V$  & $2.07$ & $0.33$ & $1.53$ & $|H-2\rightarrow L\rangle$$(0.5186)$\tabularnewline
 &  &  &  & $|H\rightarrow L+1;H\rightarrow L\rangle$$(0.3222)$\tabularnewline
\hline 
$VI$  & $2.32$ & $1.78$ & $1.05$ & $|H-3\rightarrow L\rangle$$(0.5462)$\tabularnewline
 &  &  &  & $|H-1\rightarrow L;H-1\rightarrow L+2\rangle$$(0.2991)$\tabularnewline
 &  &  &  & $|H\rightarrow L+2\rangle$$(0.2941)$\tabularnewline
\hline 
$VII$ & $2.85$ & $0.88$ & $0.60$ & $|H-9\rightarrow L\rangle$$(0.3857)$\tabularnewline
 &  &  &  & $|H-1\rightarrow L+3\rangle$$(0.3838)$\tabularnewline
\hline 
$VIII$  & $3.39$ & $0.10$ & $0.40$ & $|H-10\rightarrow L\rangle$$(0.5555)$\tabularnewline
 &  &  &  & $|H-5\rightarrow L;H\rightarrow L\rangle$$(0.3884)$\tabularnewline
\hline 
\end{tabular}
\end{table}

On examining the calculated spectra of the GQDs and quantitative information
about the corresponding excited states (Tables \ref{tab:Excited-states-contributing-GQD-64}-\ref{tab:Excited-states-contributing-SW2-GQD-64}),
we observe the following trends: (i) The first peak of the TB and
PPP-HF spectra for all these three systems is due to single excitation
of an electron from $H$ to $L$\textcolor{black}{{} level, denoted
by $|H\rightarrow L\rangle$}, corresponding to the optical gap. At
the CI level, the excited state wave function of the first peak of
GQD-64 corresponding to the optical band-gap \textcolor{black}{(Table}
\ref{tab:Excited-states-contributing-GQD-64}) is also dominated \textcolor{black}{by
the $|H\rightarrow L\rangle$} configuration. On the other hand, the
wave functions of the first peaks of SW1-GQD-64 and SW2-GQD-64 are
dominated by the transitions\textcolor{black}{{} $|H\rightarrow L+1\rangle$
and $|H-1\rightarrow L\rangle$, respectively, }in contrast to the
independent-particle predictions\textcolor{black}{. Further, the configuration
$|H\rightarrow L\rangle$ makes major contribution to the wave functions
of the excited states giving rise to the second peak of both these
SW-defect configurations (Tables \ref{tab:Excited-states-contributing-SW1-GQD-64}-\ref{tab:Excited-states-contributing-SW2-GQD-64}).
This implies that in the defective GQDs, the electron correlation
effects and the broken particle-hole symmetry due to the presence
of SW-type reconstructions, are responsible for the appearance of
these first peaks below the ones corresponding to the $|H\rightarrow L\rangle$
dominated second peaks.} (ii) At TB and PPP-HF levels, \textcolor{black}{introduction
of} SW defects at the edge/core of GQD-64 results in marginal blue-shift/no
shift of the optical gap, as compared to GQD-64. In contradiction,
the PPP-CI calculations predict redshifts in these optical gaps compared
to GQD-64, with the decrease being more pronounced for SW1-GQD-64.
Further, the entire PPP-CI absorption spectrum of SW2-GQD-64 exhibits
a red-shift in comparison to GQD-64, while no such trend is discernable
for SW1-GQD-64. \textcolor{black}{(iii) The intensities of the absorption
peaks corresponding to the $|H\rightarrow L\rangle$ excitation obtained
from TB, PPP-HF and PPP-CI approaches, decrease drastically with the
introduction of }SW-defects. \textcolor{black}{In addition, the relative
strengths of the defect-induced first peaks for SW1-GQD-64/SW2-GQD-64
at the PPP-CI level are always less as compared with the $|H\rightarrow L\rangle$
peaks of the spectra. }(iv) The first PPP-CI peaks are not the most
intense ones for all the three GQDs, in complete contradiction to
the results obtained from the independent particle model, signifying
the \textcolor{black}{importance of electron correlation effects}.
(v) The incorporation of SW-type bond rotations at the edge/middle
of the quantum dot leads to a significant blue/red-shift of the maximum
intensity (MI) peak in comparison to that of GQD-64 at CI level, which
is in contrast to the marginal/no-shift of this peak seen in the TB
and PPP-HF results. (vi) The first (I) and sixth (VI) high \textcolor{black}{intensity}
absorption peaks of pristine GQD-64 are in the infra-red and the UV
ranges, respectively, while rest of the strong absorption peaks are
in the visible range, resulting in 59\% of the total absorption spectrum
being in the visible domain. When SW-type bond rotations are present
at the edge of GQD-64, all strong absorption peaks except the tenth
(X) high \textcolor{black}{intensity} peak is in the visible spectrum,
increasing the area under visible range to 65\%. Also, when SW defects
are created at the core of GQD-64, all the high \textcolor{black}{intensity}
peaks are in the visible range, resulting in 83\% of the total CI
absorption spectrum being in that domain. These results highlight
that the introduction of SW defects in GQD-64 shifts the strong absorption
peaks to the visible range, leading to substantial enhancement of
visible light absorption. In addition, this improvement in visible
light absorption is more pronounced for the SW reconstructions located
at the core, as compared to those at the edge of pristine GQD-64.\textcolor{red}{{}
}\textcolor{black}{(vii) }The strength of the MI peaks computed by
the PPP-CI methodology are enhanced/diminished when SW defects are
at the edge/centre of the quantum dots, as compared to that of the
pristine GQD-64. But, the corresponding peak intensities for the defective
GQDs are reduced as compared to GQD-64, when computed using the independent-particle
approaches. (viii) The wave functions of the excited states responsible
for absorption peaks in GQDs with SW defects demonstrate stronger
mixing of several single excitations along with significant contribution
from doubly-excited configurations, emphasizing the importance of
electron correlation effect in these systems, as compared to pristine
GQD-64. (ix) The number of absorption peaks obtained from all the
three methodologies increases when SW-type reconstructions are present
in the system. 

We now provide a detailed discussion of the CI results for the absorption
spectra of individual configurations. Figure \textcolor{black}{\ref{fig:GQD-64-mrsdci}}\textcolor{red}{{}
}depicts the computed linear absorption spectrum of GQD-64. The first
peak at 1.15 eV (near IR range), corresponding to the optical band-gap,
is quite intense and exhibits both x as well as y - polarization.
However, the magnitude of transition dipole moment along y-axis is
significantly higher than its value along the x-axis. The most intense
peak (peak III at energy 2.62 eV) with mixed x-y polarization is due
to two closely spaced excited states having energies 2.58 eV and 2.67
eV (blue range of visible spectrum), respectively. The wave-functions
of these excited states have dominant contributions from the configurations
$|H-1\rightarrow L+1\rangle$ and $|H-3\rightarrow L\rangle+c.c.$,
respectively, where c.c. denotes the charge conjugated configuration.
The other two high intense peaks (peak V and peak VI) of the absorption
spectrum at energies 3.25 eV (violet range of visible spectrum) and
3.38 eV (near UV) are primarily due to single excitations involving
higher energy levels $|H-5\rightarrow L\rangle+c.c.$ and $|H\rightarrow L+7\rangle-c.c.$,
respectively. Peak V exhibits both x as well as y - polarization while
peak VI is dominantly polarized along x-axis. Hence, our results indicate
that single excitations are mainly responsible for the appearance
of absorption peaks. Further, in contrast to the results obtained
from the independent particle model, most of the intense peaks occur
at the high energy end of the spectrum, signifying the importance
of inclusion of electron correlation effects in predicting the absorption
peak profile. 

In case of SW1-GQD-64 (fig. \ref{fig:SW1-GQD-64-mrsdci}), the first
weak peak (0.57 eV) with almost equal polarization along both x- and
y-axis is primarily due to the single excitation $|H\rightarrow L+1\rangle$,
with partial contribution from double excitation $|H\rightarrow L;H\rightarrow L+1\rangle$.
The low intensity second peak (0.87 eV) defining the optical band-gap
also derives remarkable contribution from the double excitation $|H\rightarrow L;H\rightarrow L\rangle$,
\textcolor{black}{and exhibits higher intensities for photons polarized
along the }\textbf{\textcolor{black}{$y$}}\textcolor{black}{-axis,
as compared to those polarized along the $x$-axis.} The wave-function
of the excited state giving rise to the most intense peak (peak IX
at 3.11 eV, corresponding to the violet range of the visible spectrum)
with mixed $x-y$ polarization is largely composed of the singly excited
configurations $|H-1\rightarrow L+2\rangle$ and $|H-5\rightarrow L\rangle$.
In addition, the wave-function of the excited state leading to the
appearance of high \textcolor{black}{intensity} peak VIII at 2.91
eV (violet range), with a mixed $x-y$ character, manifests a strong
mixing of singly and doubly excited configurations $|H-4\rightarrow L\rangle$,
$|H\rightarrow L+6\rangle$, $|H\rightarrow L+4\rangle$ and $|H\rightarrow L;H-1\rightarrow L+1\rangle$,
with almost equal contributions. Similarly, the excited state responsible
for the high \textcolor{black}{intensity} peak X at energy 3.38 eV
(near UV) having mixed $x-y$ polarization \textcolor{black}{is due
to single excitations $|H\rightarrow L+11\rangle$ and $|H-7\rightarrow L\rangle$,
which involve orbitals far away from the fermi level. }

For SW2-GQD-64 (fig. \ref{fig:SW2-GQD-64-mrsdci}), the first two
absorption peaks have very low intensities. The first peak (0.65 eV)
with higher magnitude of transition dipole moment along x-axis, as
compared to y-axis, is largely composed of single excitation $|H-1\rightarrow L\rangle$,
with a weak contribution from $|H\rightarrow L\rangle$. The second
peak (1.05 eV) with mixed $x-y$ character is mainly due to the excitation
$|H\rightarrow L\rangle$, with some contribution from $|H-1\rightarrow L\rangle$
configuration. The $x-y$ polarized most intense peak $\text{VI}$
at 2.32 eV (green spectral line) is mainly due to the single and double
excitations $|H-3\rightarrow L\rangle$, $|H\rightarrow L+2\rangle$
and $|H-1\rightarrow L;H-1\rightarrow L+2\rangle$. The high intensity
peaks IV (1.86 eV - red spectral line) and VII (2.85 eV - blue spectral
line) with mixed $x-y$ character exhibits strong mixing of equally
contributing configurations $|H-1\rightarrow L+2\rangle$, $|H-1\rightarrow L+1;H\rightarrow L\rangle$,
$|H-1\rightarrow L+1\rangle$ and $|H-9\rightarrow L\rangle$, $|H-1\rightarrow L+3\rangle$,
respectively. Furthermore, the excited state giving rise to high intensity
peak V at 2.07 eV (orange range of visible spectrum) with larger polarization
along y-axis, as compared to x-axis, is mainly composed of the single
and double excitations $|H-2\rightarrow L\rangle$ and $|H\rightarrow L+1;H\rightarrow L\rangle$.
The contribution of these double excitations in the absorption spectrum
is strictly due to the presence of strong electron correlation effects
in SW-defected systems.

\section{Conclusions\label{sec:Conclusions}}

In conclusion, we have performed very large-scale electron-correlated
computations employing the PPP Hamiltonian to critically analyze the
role played by SW-type bond rotations present at either the zigzag
edge, or at the core of GQD-64 in determining its structural, electronic
and optical properties. Our computations indicate that SW-type defects
increases the energy of pristine GQD-64. In addition, the presence
of SW defects at the zigzag edge leads to a lower energy than its
location at the middle of the quantum dot. Our electron-correlated
results demonstrate that the introduction of SW-type defects is responsible
for the appearance of defect-induced peak below the optical band-gap.
Also, the variation of the entire absorption peak profile is critically
dependent upon the location of SW-type reconstruction. In addition,
our studies emphasize that electron correlation effects become more
dominant for SW-defected GQDs. Finally, our results signify that the
introduction of SW defects in a \textcolor{black}{systematic} manner
(edge/core of GQD-64) shifts the high intensity edge of optical absorption
spectrum to the visible range, leading to significant enhancement
of visible light absorption. This mechanism can be efficiently exploited
to fabricate novel devices for light harvesting and optoelectronics. 

\bibliographystyle{apsrev4-1}
\addcontentsline{toc}{section}{\refname}\bibliography{Reczag_SW}

\end{document}